\documentclass[
twocolumn,
aps,
superscriptaddress, 
amsmath,
amssymb,
amsfonts,
floatfix,
square
]{revtex4-1}
\usepackage[colorlinks=true, allcolors=blue]{hyperref}
\usepackage{bm}
\usepackage{lipsum}
\usepackage[utf8]{inputenc}
\usepackage[nolist]{acronym}
\usepackage[caption=false]{subfig}
\usepackage{graphicx}
\usepackage{siunitx}
\usepackage{import}
\usepackage{float}
\usepackage[dvipsnames]{xcolor}
\usepackage{chemformula}
\setcitestyle{numbers,square}

\begin{document}
\title{Zero temperature superconductor - edge metal - insulator transition in \(2D\) bosonic systems}
\author{H\aa vard H. Haugen} 
\affiliation{\footnotesize Department of Physics, Norwegian University of Science and Technology, NO-7491, Trondheim, Norway}
\affiliation{\footnotesize Center for Quantum Spintronics, Department of Physics, Norwegian University of Science and Technology, NO-7491, Trondheim, Norway}
\author{Asle Sudb\o}
\affiliation{\footnotesize Department of Physics, Norwegian University of Science and Technology, NO-7491, Trondheim, Norway}
\affiliation{\footnotesize Center for Quantum Spintronics, Department of Physics, Norwegian University of Science and Technology, NO-7491, Trondheim, Norway}
\date{\today}

\begin{abstract}
    Motivated by the recent experimental observation of an intermediate bosonic metallic state in the two-dimensional superconductor-insulator transition at \(T=0\), we study an extended Bose Hubbard model in the limit of large number of particles per site. Using a representation of this in terms of two coupled $XY$ models, we find, in addition to an insulating phase and a $(2+1)D$ superfluid phase, two other phases. One phase is a $2D$ superfluid phase where a crossover from $(2+1)D$ to $2D$ has taken place as a result of incipient charge ordering, signalled by $\theta$ disordering, and which is closely related to a supersolid phase. The other new phase is an edge metal state characterized by zero superfluid stiffness, zero charge ordering, and zero bulk compressibility. However, the edge compressibility of the system is nonzero. While we do not find any intermediate state with $2D$ metallic conductivity, we are able to connect these results to STM experiments on \ch{MoS2} showing brims of finite density of states around the entire edge of $2D$ \ch{MoS2} samples.         
\end{abstract}
\maketitle

\section{Introduction}
\begin{acronym}
  \acro{SC}{superconductor}
  \acro{M}{metal}
  \acro{EM}{edge metal}
  \acro{I}{insulator}
  \acro{BKT}{Berezinskii–Kosterlitz–Thouless}
\end{acronym}
The zero temperature phases of two-dimensional systems has been a topic of intense research for many decades. Early works showed that even a small amount of disorder is enough to localize all electronic states, implying that electrons can not form a metallic phase at \(T=0\) \cite{PhysRevLett.42.673}.  Hence, the zero temperature ground state of two-dimensional systems should either be superconducting (superfluid) or insulating. However, experimental evidence of a metallic state at ultra-low temperatures at the phase-boundary between the superconducting and insulating phases has been reported \cite{PhysRevB.40.182}. This was explained using the idea that even outside the superconducting phase the relevant degrees of freedom are still bosonic, in the form of preformed or phase-incoherent Cooper pairs. The analysis of these bosonic models led to the prediction that the resistance is universal at the \ac{SC}-\ac{I} transition \cite{PhysRevLett.64.587, PhysRevB.44.6883, PhysRevB.49.12115, PhysRevB.49.9794}. 

More recent experiments have shown that the metallic phase is not necessarily a point in parameter-space, but can exist for a wide parameter regime, typically intervening the superconducting and insulating phases. This metallic behaviour has been found in a number of experimental setups, and the phase transition from \ac{SC}/\ac{I} to a metallic state can be induced by varying different experimental parameters \cite{RevModPhys.91.011002}. A magnetic field driven transition has been observed in \ch{MoGe}, \ch{Ta}, \ch{TaN} and \ch{InO} thin films \cite{PhysRevLett.76.1529, PhysRevLett.82.5341, PhysRevB.64.060504, PhysRevB.73.100505, doi:10.1126/sciadv.1700612}. The metallic phase has also been observed in granular systems, where a gate voltage is used to tune the effective Josephson coupling between superconducting grains \cite{PhysRevLett.78.2632, han2014collapse, bottcher2018superconducting}. Similar results with gate-voltage tuned transitions have been obtained for homogeneous systems \cite{doi:10.1126/science.1259440, doi:10.1126/science.1228006, chen2018carrier}. The final category of experiments include systems where the geometry is tuned. Metallic behaviour has been observed both in homogeneous systems where the film thickness is tuned, and in granular systems where the distance between the superconducting grains is tuned \cite{PhysRevB.40.182, eley2012approaching}. A particularly illuminating study of YBCO \cite{doi:10.1126/science.aax5798} used reactive ion etching to tune the thickness and crystalline properties of superconducting islands. Measurements of the magneto-conductance oscillation period were consistent with \(2e\) charge carriers, which demonstrates that Cooper pairs participate in transport in the metallic state.

Despite the plethora of experimental evidence for a zero temperature metallic phase, consensus is that no satisfactory theoretical description of the \emph{anomalous metal} or \emph{Bose metal} exists to date \cite{RevModPhys.91.011002}. The fact that it can be found in a wide variety of materials points to some fundamental effect that is missing in our current understanding of these systems. Considerable theoretical effort has gone into the study of systems where the superconducting order parameter is coupled to a dissipative heat bath in the context of resistively shunted Josephson junctions \cite{PhysRevLett.56.2303, PhysRevB.37.3283, PhysRevLett.79.2730, PhysRevB.63.125322, PhysRevB.72.060505, PhysRevB.85.224531, PhysRevB.91.205129}. Another explored avenue is the glassy phases found in Josephson junction array models with local disorder introduced through random Josephson couplings or gauge field \cite{PhysRevB.45.10490, PhysRevLett.89.027001, doi:10.1126/science.1088253, PhysRevB.102.184503, granato2021magnetic}. In Ref.  \cite{PhysRevB.60.1261}, it was proposed that the \ac{SC}-\ac{M}-\ac{I} transition should be viewed as a two order parameter problem, one describing superconducting ordering and the other describing charge ordering. This allows for a scenario where 
an intermediate phase exists where
superconducting phase coherence is destroyed, but charge order is not yet established.  
Mapping the system onto bosons interacting strongly with a gauge-field and comparing with previously obtained results \cite{PhysRevB.48.16641}, Ref. \cite{PhysRevB.60.1261} suggested that  a Bose metal phase containing gapless diffusive charge excitations in the bulk spectrum would result. 
This was then proposed to constitute the zero-temperature Bose metal.

In this paper, we analyze the same model as in Ref. \cite{PhysRevB.60.1261} model using large-scale Monte Carlo simulations. Our results confirm that an intermediate phase exists, but with a slightly different physical interpretation than proposed in Ref.~\cite{PhysRevB.60.1261}. We find that the bulk compressibility vanishes in the intermediate phase, implying that charge excitations are gapped in the bulk. We show that the order parameter of the intermediate phase instead corresponds to a non-zero edge compressibility. From this, we conclude that the intermediate phase is a bulk insulator with conducting edge states. Throughout the paper we will refer to this phase as a \ac{EM}. We also find a fourth critical superconducting (CSC) phase, with \(2D\) superfluid ordering but no order along the temporal direction, as opposed to the regular SC phase. We present a detailed study of the different possible phase transitions and map out a phase diagram for the model. We comment on the differences between the phase diagram obtained here and in Ref.~\cite{PhysRevB.60.1261}.   


\section{Model}\label{sec:Model}
We start by considering the following quantum rotor model on a \(L\times L\) two-dimensional lattice 
\begin{alignat}{1}
     H = & -J\sum_{i, \alpha} \cos(\phi_{i+\alpha}-\phi_{i}) - \bar{\mu}\sum_{i}\hat{m}_i \label{eq:EBHM_Hamiltonian}\\
     & + \frac{U}{2}\sum_i (\hat{m}_i)^2 + V\sum_{i,\alpha}\hat{m}_i \hat{m}_{i+\alpha} \nonumber
\end{alignat}
where \(\alpha = \hat{x}, \hat{y}\). Each site has an associated phase \(\phi_i\) and number fluctuation operator \(\hat{m}_i = \hat{n}_i - n_0\), where \(\hat{n}_i\) counts the number of bosons at site \(i\) and \(n_0\) is the average boson density for $\bar{\mu} =0$. The phase and number operators have a non-trivial commutation relation
\begin{equation}
    [\hat{m}_i, \hat{\phi}_j] = i\delta_{ij}.
\end{equation}
The first term in Eq.~\eqref{eq:EBHM_Hamiltonian} describes the kinetic energy of the bosons. The second term is the chemical potential energy of the system, where \(\bar{\mu} = \mu + U/2\) is a renormalized chemical potential. In all the results presented in this paper we have set \(\bar{\mu} = 0\), which implies \(\sum_i \langle m_i \rangle = 0\). The third and fourth term represents the on-site and nearest neighbor potential energy respectively. Eq.~\eqref{eq:EBHM_Hamiltonian} describes the physics of the extended Bose Hubbard model in the limit of large integer boson density \(n_0\) \cite{PhysRevB.40.546}. For a granular superconducting film, the parameters of the model can be determined experimentally from the relation \cite{PhysRevB.60.1261, PhysRevB.39.6441}
\begin{equation}
    J = \frac{R_Q}{2R_n}\Delta_0,
\end{equation}
where \(R_Q = \sigma_Q^{-1} = h/(2e)^2\), \(R_n\) is the normal state resistance and \(\Delta_0\) is the pairing gap. \(U\) and \(V\) are given by the inverse of the capacitance matrix \(C_{\alpha\beta}\) of the superconducting grains. 

In Ref.~\cite{PhysRevB.60.1261}, it was shown that the quantum rotor model introduced in Eq.~\eqref{eq:EBHM_Hamiltonian} can be transformed into two coupled \(XY\) models in 2+1 dimensions. Their derivation is summarized here for completeness. We start by writing the model introduced in Eq.~\eqref{eq:EBHM_Hamiltonian} on the following form
\begin{alignat}{1}
    H = & -J\sum_{i, \alpha} \cos(\phi_{i+\alpha}-\phi_{i}) - \bar{\mu}\sum_{i}\hat{m}_i \label{eq:JJA_Hamiltonian}\\
    & +V_0\sum_{i}(\hat{m}_{i})^{2} + V_1 \sum_{i,\alpha}(\hat{m}_{i} + \hat{m}_{i+\alpha})^2,  \nonumber
\end{alignat}
where we have re-organized the potential terms by setting \(U/2 = V_0 + 4V_1\) and \(V=2V_1\). This is a convenient choice for the transformations done later on, and we will be discussing the model in terms of \(V_0\) and \(V_1\) in the following. Next, the model in Eq.~\eqref{eq:JJA_Hamiltonian} is written in the path integral representation, with a partition function given by
\begin{equation}
    Z = \sum_{\{m_i\}}\int_0^{2\pi} \mathcal{D}\phi e^{-S}
\end{equation}
\begin{alignat}{1}
    S =& i\sum_i m_i (\nabla_\tau \phi_i) -J\sum_{i, \alpha} \cos(\nabla_\alpha \phi_i) \label{eq:JJA_path_integral} \\
    & +V_0\sum_{i}(m_{i})^{2} + V_1 \sum_{i,\alpha}(m_{i} + m_{i+\alpha})^2, \nonumber
\end{alignat}
where we have set \(\bar{\mu}= 0\). The sum over \(i\) now runs over 2+1 dimensions, where the last dimension \(\tau\) is discrete imaginary time with periodic boundary conditions. Discretizing imaginary time introduces a length \(\Delta \tau\) between each time (Trotter) slice, which has been set to unity. The index \(\alpha \in \{\hat{x}, \hat{y}\}\) refers to spatial directions, and \(\nabla_\mu f_i = f_{i+\mu} -f_i\) is the finite difference operator. In the path integral formalism, the fields \(m\) and \(\theta\) are real numbers, not operators, and their commutation relation is taken into account by the first term in Eq.~\eqref{eq:JJA_path_integral}. Invoking the transformation \(m_i \to e^{-iQ\cdot r_i} m_i\), where \(Q = (\pi, \pi)\), the action in Eq.~\eqref{eq:JJA_path_integral}  reads
\begin{alignat}{1}
    S =& i\sum_i e^{iQ\cdot r_i}m_i (\nabla_\tau \phi_i) -J\sum_{i, \alpha} \cos(\nabla_\alpha \phi_i) \label{eq:JJA_path_integral_2} \\
    & +V_0\sum_{i}(m_{i})^{2} + V_1 \sum_{i,\alpha}(\nabla_\alpha m_i)^2. \nonumber
\end{alignat}

The next step is to Hubbard-Stratonovich decouple the \(V_1\) term, thereby introducing a new field \(p_{i\alpha}\)
\begin{equation}\label{eq:HS_decoupling_V1}
    e^{-V_1\sum_{i,\alpha}(\nabla_\alpha m_i)^2} = \int \mathcal{D}p e^{-(1/4V_1)\sum_{i,\alpha} p_{i\alpha}^2 - i\sum_{i,\alpha}m_i (\nabla_\alpha p_{i,\alpha})}.
\end{equation}
In the last term, the gradient has been shifted onto the \(p_{i,\alpha}\) factor by a partial integration. This transformation is only possible for positive values of \(V_1\), which will put a restriction on the parameter regime for which the transformed model is valid. In the original formulation, Eq.~\eqref{eq:EBHM_Hamiltonian}, a positive \(V_1\) corresponds to a parameter regime where the onsite interaction \(U\) is larger than \(2Vz\), \(z\) being the number of nearest neighbors. We can rewrite the Hubbard-Stratonich field as \(p_{i,\alpha} = \tilde{p}_{i,\alpha} + 2\pi l_{i,\alpha}\), where \(\tilde{p}_{i,\alpha} \in [0, 2\pi)\) and \(l_{i,\alpha} \in \mathbb Z\). Furthermore, the phase variable can be split into a curl and a gradient part, \(\tilde{p}_{i,\alpha} = \nabla_\alpha \theta_i + (\nabla \times \Theta_i)_\alpha\). Inserting this into Eq.~\eqref{eq:HS_decoupling_V1}, the curl contribution to \(\tilde{p}\) can be integrated out since only the divergence of \(\tilde{p}\) couples to \(m\). Furthermore, the coupling term is invariant under shifts of \(2\pi\). Thus, we obtain Eq.~\eqref{eq:HS_decoupling_V1} on the form
\begin{alignat}{1}
    \sum_{\{l_{i,\alpha}\}} & \int_0^{2\pi} \mathcal{D}\theta e^{-(1/4V_1)\sum_{i,\alpha} (\nabla_\alpha\theta_i - 2\pi l_{i,\alpha})^2 - i\sum_{i}m_i (\nabla^2_{||} \theta_i)} \nonumber \\
    & \simeq \int_0^{2\pi} \mathcal{D}\theta e^{(1/2V_1)\sum_{i,\alpha}\cos(\nabla_\alpha\theta_i) - i\sum_i m_i (\nabla^2_{||} \theta_i)}, \nonumber
\end{alignat}
where the sum over \(l\) has been executed using the Villain transformation. We also introduced the notation \(\nabla^2_{||}\) to explicitly show that the Laplace operator only includes spatial compontents. Finally, we may carry out the sum over \(m\). Collecting all the terms in the action that include \(m\) we find
\begin{alignat}{1}
    \sum_{\{m_i\}} & e^{-V_0 \sum_i m_i^2 -i\sum_i m_i (e^{-iQ\cdot r_i} \nabla_\tau \phi_i + \nabla_{||}^2 \theta_i)} \\
    & \simeq e^{(1/2V_0) \sum_i(e^{-iQ\cdot r_i} \nabla_\tau \phi_i + \nabla_{||}^2 \theta_i)},
\end{alignat}
where the Villain transformation has been used again. Combining all this leads to the following partition function for two coupled \(XY\) models
\begin{equation}
    Z = \int_0^{2\pi}\mathcal{D}\phi\mathcal{D}\theta e^{-S},
\end{equation}
\begin{alignat}{1}
    S = &-J\sum_{i,\alpha}\cos(\nabla_\alpha \phi_i) -\frac{1}{2V_1}\sum_{i,\alpha} \cos(\nabla_\alpha\theta_i)\nonumber\\
    &-\frac{1}{2V_0}\sum_{i}\cos(e^{-iQ\cdot r_i}\nabla_\tau \phi_i + \nabla_{||}^2\theta_i),\label{eq:model_charge_picture}
\end{alignat}
where the sum over \(i\) now runs over \(N = L\times L \times L_\tau\) sites in 2+1 dimensions. The action is dimensionless by absorbing the discrete imaginary time length \(\Delta \tau\) into the coupling constants.
\begin{figure}[htb]    
        \subfloat[\label{fig:phase_diagram_a}]{
        \includegraphics[width = 0.49\textwidth]{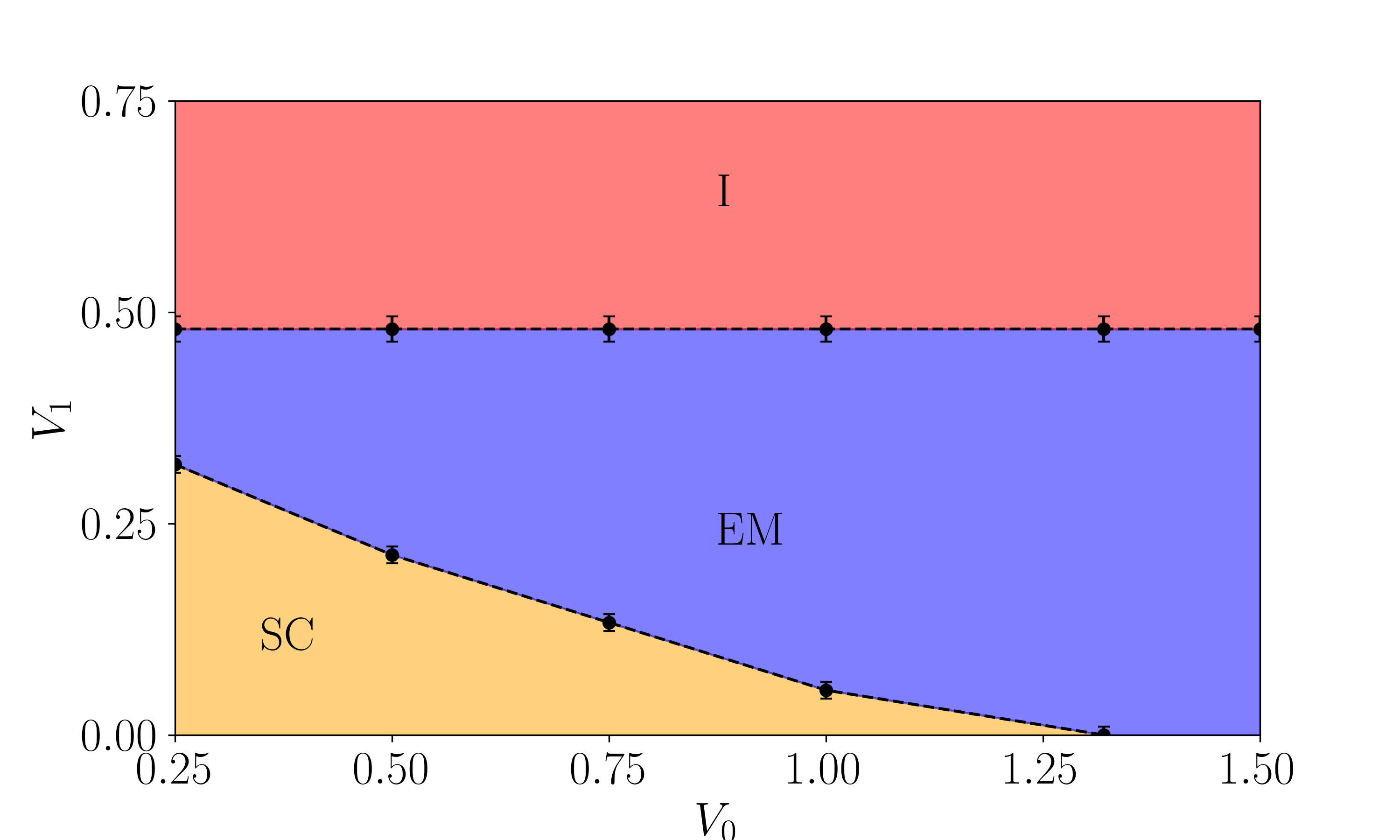}
    } \\[-3.5ex]
    
    \subfloat[\label{fig:phase_diagram_b}]{
        \includegraphics[width = 0.49\textwidth]{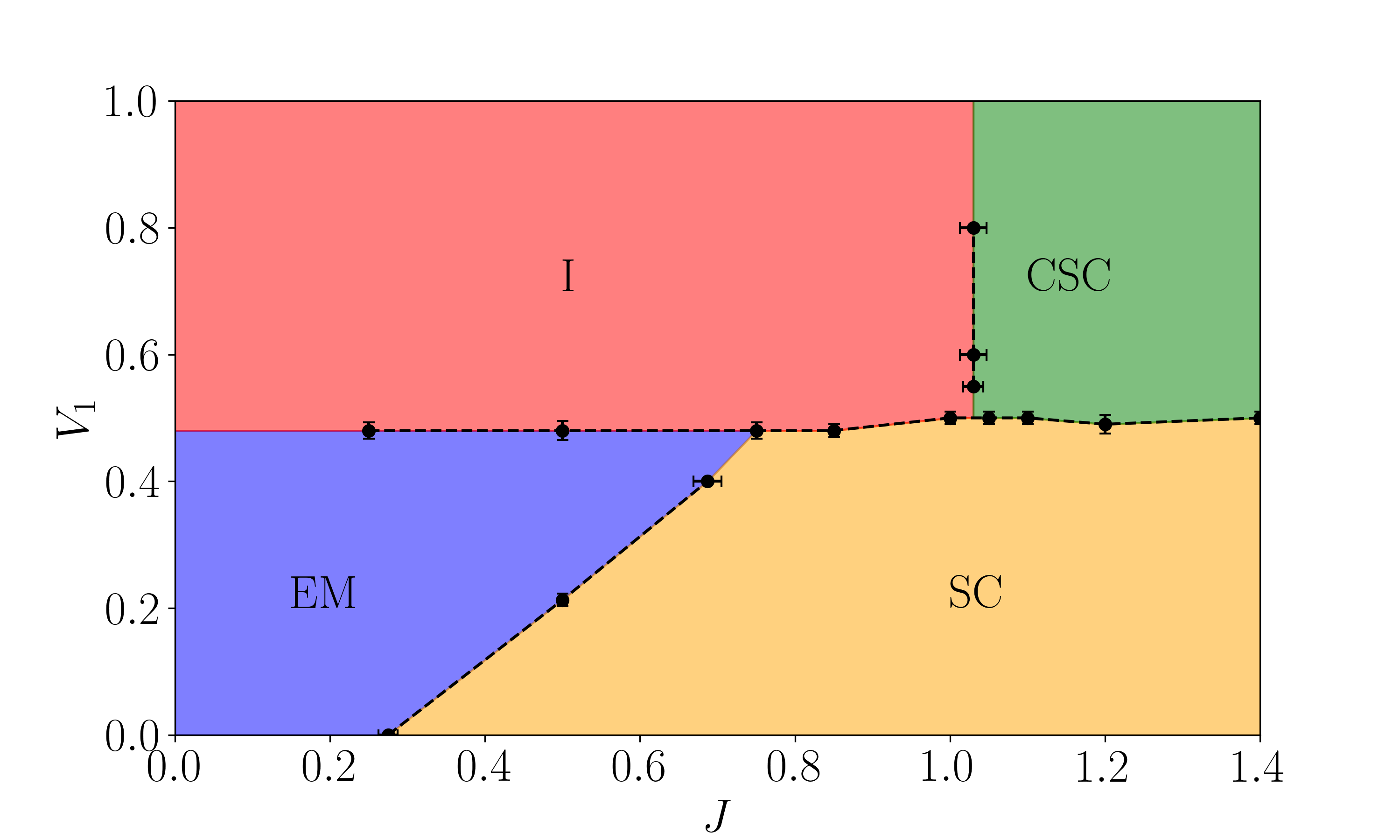}
    }
    
    \caption{Zero temperature phase diagrams for the coupled \(XY\) model in Eq.~\eqref{eq:model_charge_picture}. (a) \(V_1\)-\(V_0\) plane with \(J=0.5\) fixed. (b) \(V_1\)-\(J\) plane with \(V_0=0.5\) fixed. (a) features three distinct phases; a superconducting (SC) phase for low values of \(V_1\) and \(V_0\),  the edge metal (EM) phase characterized by non-zero edge compressibility for intermediate values of \(V_1\) and an insulating (I) phase for high values of \(V_1\). The edge-compressibility is equal to the susceptibility for creating excess charge of equal sign  on two neighboring lattice sites. (b) features four phases, namely the three phases mentioned above and in addition a strictly \(2D\) critical superfluid phase (CSC). The transition from \(3D\) to \(2D\) SC for large $J$ is driven by the proliferation of \(2D\) vortices (instantons) in the phase field $\theta$, where $\nabla \theta$ is a conjugate variable to the excess charge on two neighboring points. These results are
    obtained from Monte Carlo simulations, described in more detail in Sec.~\ref{sec:Monte_Carlo_methods}, on a \(20\times 20\times 20\) system.}
    \label{fig:phase_diagram}
\end{figure}

Key results in this paper rely on understanding the equivalence between Eq.~\eqref{eq:JJA_Hamiltonian} and Eq.~\eqref{eq:model_charge_picture}, since we use the quantum rotor model to interpret the numerical results obtained from the coupled \(XY\) model. The coupled \(XY\) model features two phase degrees of freedom, \(\phi\) and \(\theta\). Ordering of these phases is described by their phase-stiffness or helicity moduli, which is non-zero in the ordered phase. Spatial phase coherence of \(\phi\) is associated with superconductivity, i.e. the collective ordering of the superconducting phase on each grain in the system to form a superconducting state with global phase coherence. Temporal phase coherence in \(\phi\) is associated with a non-zero bulk compressibility, which measures the change in charge density in response to an infinitesimal shift of the chemical potential. Zero compressibility indicates that particle-hole excitations are gapped, yielding an insulating state, while a non-zero compressibility indicates a metallic or superconducting state. The interpretation of the field \(\theta\) is not as straightforward as the interpretation of the field \( \phi \). As Eq.~\eqref{eq:HS_decoupling_V1} (and onward) shows, the field \(\theta\) arises from decoupling a term on the form \((m_i + m_{i+\alpha})^2\) in the un-rotated basis. To be precise, \(p \propto \nabla\theta\) is the dual variable to \(m_i + m_{i+\alpha}\), so ordering of \(\theta\) is tied to disordering the field \(m_i + m_{i+\alpha}\). In Sec.~\ref{sec:Observables}, we show explicitly that the order parameter of \(\theta\) is closely linked to a susceptibility to induce excess charge on two neighboring sites so that when \(\theta\) is disordered charge density wave is the only possible charge excitation. However, due to the transformation in Eq.~\eqref{eq:HS_decoupling_V1}, the coupled XY model is only valid in a parameter regime where the nearest neighbor interaction is too weak to establish charge density wave ordering, and consequently we find a uniform charge distribution when \(\theta\) is both ordered and disordered. There is still an observable difference between these two cases in the valid parameter regime. In Appendix \ref{ap:Edge_compressibility} we show that spatial phase coherence in \(\theta\) corresponds to a non-zero edge compressiblity, that will be defined in Sec.~\ref{sec:Observables}. Like the bulk compressibility, we can use the edge compressibility to determine whether the edge states feature gapless excitations or not. The edge compressibility is non-zero in the phase where \(\theta\) is ordered, which corresponds to the phase where both uniform and charge density wave excitations are possible in the bulk. In the phase where \(\theta\) is disordered, and only charge density wave excitations are possible in the bulk, the edges are incompressible. 

Among the different combinations of ordering in the two fields we find that all four are realized in the parameter regime considered, summarized in Table \ref{tab:Phases_ordering}. \emph{i) Spatial and temporal ordering in \(\phi\), spatial ordering in \(\theta\)}. This corresponds to a compressible superconducting (SC) phase with non-zero superfluid stiffness. \emph{ii) Spatial and temporal disordering in \(\phi\),  spatial ordering in \(\theta\)}. This corresponds to a phase with zero superfluid stiffness and an incompressible bulk. The edge compressibility remains non-zero, yielding an edge metal (EM) phase with non-zero edge conductance. \emph{iii) \(\phi\) and \(\theta\) disordered.} This corresponds to a state with zero superfluid stiffness along with incompressible bulk and edges yielding a fully insulating (I) phase. \emph{iv)  Spatial ordering and temporal disordering in \(\phi\), \(\theta\) disordered}. This state is realized above a critical value of $J > J_c \approx 1.03$, and corresponds to a critical superconducting (CSC) state with finite superfluid stiffness and zero bulk as well as edge compressibility. It is closely related to the critical superconducting state found in previous works using quantum Monte Carlo simulations on $(2+1)D$ quantum rotor models with Caldeira-Legget dissipation \cite{PhysRevB.85.224531,PhysRevB.91.205129}. However, the crossover/transition from $(2+1)D$ to $2D$ superfluidity is not driven by varying an explicit dissipation strength as in Refs.~\cite{PhysRevB.85.224531,PhysRevB.91.205129}, but by the disordering of $\theta$, equivalently incipient charge ordering. It is therefore closely related to a supersolid phase. An overview of \((V_1, V_0)\)- and \((V_1, J)\)-parameter space is presented in the phase diagram in Fig.~\ref{fig:phase_diagram}.

\begin{table}[htb]
\caption{Summary of how the two phase fields \(\theta\) and \(\phi\) are ordered in each of the four phases found in this paper.}
\centering
\begin{tabular}{l@{\hskip 0.6in}l@{\hskip 0.6in}l}
\hline\hline
Phase & \(\phi\) ordering & \(\theta\) ordering \\
\hline
SC & Spatial and temporal & Spatial \\
EM & Disordered & Spatial \\
I & Disordered & Disordered \\
CSC & Spatial & Disordered \\
\hline\hline
\end{tabular}
\label{tab:Phases_ordering}
\end{table}

\section{Monte Carlo methods}\label{sec:Monte_Carlo_methods}
To investigate the different zero temperature phases of the coupled \(XY\) model, we need to minimize the action in Eq.~\eqref{eq:model_charge_picture}. This can be achieved using Monte Carlo simulations, where we obtain averages of various observables for a given set of system parameters through a weighted walk through the configuration space of the two fields using the Metropolis Hastings algorithm \cite{Katzgraber09, Press07, Newman99}. The coupled \(XY\) model is already discretized, and we use a forward finite difference for the gradients. The Laplace operator is discretized using a central finite difference, to avoid breaking four fold rotation symmetry, so that the action we use in numerical simulations explicitly reads
\begin{widetext}
\begin{alignat}{1}
    S = & -J\sum_{i,\alpha} \cos(\phi_{i+\alpha}-\phi_{i})-\frac{1}{2V_1}\sum_{i,\alpha}\cos(\theta_{i+\alpha}-\theta_{i}) \label{eq:model_charge_picture_disc}\\
    &-\frac{1}{2V_0}\sum_i\cos((-1)^{x_i + y_i}(\phi_{i+\tau}-\phi_i) + \theta_{i+x} + \theta_{i-x}+\theta_{i+y} + \theta_{i-y} - 4\theta_{i}), \nonumber
\end{alignat}
\end{widetext}
where \(\alpha = \hat{x},\hat{y}\) refers to in-plane directions. The system is $2+1$ dimensional so that the sum over \(i\) runs over a total of \(N=L\times L \times L_\tau\) lattice sites. The sum over imaginary time runs from zero to \(\beta\), where \(\beta\) is inverse temperature, so to find the zero temperature behaviour we should consider a system that extends to infinity in all three dimension. This is enforced through periodic boundary conditions, and we can use finite size scaling to infer the behaviour in an infinite system.

Using Eq.~\eqref{eq:model_charge_picture_disc}, the Monte Carlo simulations proceed as follows. Starting from some initial configuration, typically a state that is uniform in $(\phi_{i},\theta_{i})$, we propose an updated state where one of the fields on a single site has been changed. The value of the updated field is drawn from a uniform probability distribution over the interval \([\phi_i - \phi_m, \phi_i + \phi_m]\) (mod \(2\pi\)), where \(\phi_m\) was typically set to \(2\pi/3\). This new configuration is then accepted if
\begin{equation}
    \ln r \leq -\Delta S
\end{equation}
where \(\Delta S = S_{\mathrm{new}} - S_{\mathrm{old}}\) is the change in the action from the updated field, and \(r \in [0, 1]\) is a random number. Thus, if the action decreases the change is always accepted and if it increases the change is accepted with some probability depending on the change. This leads to a weighted walk, where high probability configurations with a smaller value of the action are visited more frequently, which simulates the zero temperature quantum fluctuations of the system. The procedure of updating a field value is done sequentially for both fields \(\phi\) and \(\theta\) for every lattice site, referred to as a Monte Carlo sweep. In simulations we do millions of such sweeps, while sampling observables after every fixed number of sweeps to obtain statistical averages. We also do an initial set of sweeps to thermalize the system in a high probability configuration before starting the measurements. Errors in the results were estimated using the jackknife method \cite{Efron_1_1979}.

\section{Observables}\label{sec:Observables}
To characterize the possible phases of the model along with the phase transitions separating them, we study a number of observables. A key observable of the \(XY\) model is the helicity modulus, or phase stiffness. The helicity modulus is given by the curvature of the free energy with respect to an infinitesimal twist in one of the phases along some direction \(\hat{\mu}\) in the system \(\phi_i \to \phi_i + \gamma_\mu \hat{\mu}\cdot r_i\). We can measure the helicity modulus of both $\phi_i$ and $\theta_i$, along spatial and temporal directions. They are given on the form
\begin{equation}\label{eq:hel_mod_generic}
    \Upsilon_{\mu}^{\Theta} = \frac{1}{N}\left[\langle \epsilon_{\mu}^{\Theta}\rangle - \langle (I_{\mu}^{\Theta})^2\rangle\right],
\end{equation}
where \(\Theta = \theta, \phi\) labels the phase field and \(\mu=x,y,\tau\) labels the direction of the twist. The brackets \(\langle ...\rangle\) denotes averaging over quantum fluctuations. We sample three different helicity moduli, where the explicit expressions to be used in Eq.~\eqref{eq:hel_mod_generic} are
\begin{equation}
     \epsilon_{\alpha}^\phi = J\sum_{i}\cos(\phi_{i+\alpha}-\phi_{i}),
\end{equation}
\begin{equation}
    I_{\alpha}^\phi = J\sum_{i}\sin(\phi_{i+\alpha}-\phi_{i}),
\end{equation}
\begin{equation}\label{eq:epsilon_phi_tau}
    \epsilon_\tau^{\phi} = \frac{1}{2V_0}\sum_i \cos(e^{-iQ\cdot r_i}(\nabla_\tau \phi_i) + \nabla^2\theta_i),
\end{equation}
\begin{equation}\label{eq:I_phi_tau}
    I_\tau^{\phi} = \frac{1}{2V_0}\sum_i e^{-iQ\cdot r_i} \sin(e^{-iQ\cdot r_i}(\nabla_\tau \phi_i) + \nabla^2\theta_i),
\end{equation}
\begin{equation}\label{eq:epsilon_theta_alpha}
    \epsilon_\alpha^\theta = \frac{1}{2V_1}\sum_{i}\cos(\theta_{i+\alpha}-\theta_{i}),
\end{equation}
\begin{equation}\label{eq:I_theta_alpha}
    I_\alpha^\theta = \frac{1}{2V_1}\sum_{i}\sin(\theta_{i+\alpha}-\theta_{i}).
\end{equation}
The helicity modulus of \(\phi\) in the spatial directions is proportional to the superfluid stiffness. We are only interested in whether this is zero or finite, so we set these equal for simplicity
\begin{equation}\label{eq:SF_stiffness}
    \Upsilon_\alpha^\phi = \rho_s,
\end{equation}
where a non-zero superfluid stiffness signals superconductivity. The helicity modulus of \(\phi\) in the temporal direction is equal to the bulk compressibility of the system
\begin{equation}\label{eq:bulk_compressibility}
    \Upsilon_\tau^\phi = \kappa \equiv \frac{\partial \langle m \rangle}{\partial \mu}.
\end{equation}
This has previously been shown for \(V_1 = 0\) in Ref.~\cite{PhysRevB.40.546}, and in Appendix \ref{ap:Bulk_compressibility} we show that this equality also holds when the \(V_1\) term is included. The insulating phase is incompressible due to an energy gap for particle-hole excitations, and we will use the compressibility to distinguish between insulating and conducting phases. Finally, the helicity modulus of \(\theta\) in the spatial directions is equal to the edge compressibility of the system which we define as
\begin{equation}\label{eq:edge_compressibility}
    \Upsilon_\alpha^\theta = \kappa_{e,\alpha} \equiv \frac{1}{N}\frac{\partial \langle M_{\alpha}\rangle}{\partial \mu_{e,\alpha}},
\end{equation}
\begin{equation}\label{eq:edge_occupation_number}
    M_\alpha = \sum_{i}{}^{'}(m_{1,i} + m_{L,i}),
\end{equation}
where the primed sum runs over the two edges of the system that are orthogonal to \(\alpha\). Note that we only differentiate with respect to the chemical potential at those same two edges. Similarly to the bulk compressibility, the edge compressibility is used to determine whether a given phase features conducting or insulating edge states. To our knowledge the edge compressibility has not been studied in similar models previously, and we provide a more detailed derivation of Eq.~\eqref{eq:edge_compressibility} in Appendix~\ref{ap:Edge_compressibility}. We would like to emphasize that the edge compressibility, like the helicity modulus, is a bulk observable which does not depend on the boundary conditions. To observe an edge current the boundary conditions would have to be altered, so what we actually measure in this paper is the bulk signal of a phase that should exhibit edge currents if we were to introduce an edge by altering the boundary conditions. To understand what ordering in \(\theta\) means for the bulk of the system, consider the following correlation function
\begin{equation}\label{eq:Correlation_function_sigma_m}
    G_{\Sigma m}^{\alpha}(r_i - r_j) = \langle \Sigma_\alpha m_i \Sigma_\alpha m_j \rangle - \langle \Sigma_\alpha m_i \rangle \langle \Sigma_\alpha m_j\rangle,
\end{equation}
with 
\begin{equation}
    \Sigma_\alpha m_i \equiv (m_i + m_{i+\alpha})e^{iQ\cdot r_i}.
\end{equation}
We can then define the following susceptibility 
\begin{equation}\label{eq:susceptibility_theta}
    \chi_{\Sigma_m}^{\alpha} = \lim_{q \to 0} \frac{1}{N^2}\sum_{ij} G_{\Sigma m}^{\alpha}(r_i - r_j) e^{iq\cdot (r_i - r_j)} \sim \Upsilon_{\alpha}^{\theta}.
\end{equation}
This is not strictly equal to the helicity modulus, as the long wavelength limit and averages over lattice and Monte Carlo sweeps are taken in different order, but they will show the same behaviour. This is shown in Appendix \ref{ap:Equivalence}. Thus, when the helicity modulus of \(\theta\) is zero the susceptibility \( \chi_{\Sigma_m}^{\alpha}\) is also zero, and excitations on the form \( (m_i + m_{i+\alpha})e^{iQ\cdot r_i}\) are frozen out. This means that the only possible excitation in \(m\) when \(\theta\) is disordered are configurations with \(m_i = - m_{i+\alpha}\), in other words charge density wave ordering. However, we find that when \(\theta\) is disordered \emph{all} fluctuations in \(m\) are suppressed and stable charge density wave ordering is never established. Therefore we will refer to this as incipient charge ordering.

Correlations in the two phase-fields can also provide information about the different phases. We will investigate the spatial correlation-function of the field \(\theta\), given by
\begin{equation}\label{eq:correlation_function_theta}
    G_\theta (r_i) = \langle \cos (\theta_i - \theta_0)\rangle ,
\end{equation}
where \(\theta_0\) is the phase at the origin of our coordinate system and \(\theta_i\) is the phase at some site \(i\) in the same time-slice. Note that this is just a convenient choice and we could measure the correlation function between any two sites as it only depends on the relative distance between them. 

For the phase \(\phi\) we consider the possibility of both in-plane and temporal ordering, through the superfluid density and bulk compressibility respectively. We can also measure the simultaneous ordering in all directions through the \(3D\) \(XY\) order parameter, which is given by
\begin{equation}\label{eq:OP_phi}
    |m^{\phi}| = \frac{1}{N}\left\langle\left|\left(\sum_{i}\cos(\phi_i), \sum_{i}\sin(\phi_i)\right)\right|\right\rangle.
\end{equation}
This order parameter is equivalent to the magnetization for magnetic systems, and 2+1 D phase coherence is characterized by a non-zero value of \(|m^{\phi}|\). We note that the absolute value has to be taken \emph{before} the Monte Carlo average, because the direction of the order parameter in a fully correlated state can change over a number of Monte Carlo sweeps. We can also determine the critical coupling of a \(3D\) \(XY\) phase transition by analyzing the Binder cumulant of the order parameter
\begin{equation}\label{eq:Binder_cumulant}
    U_L(m) = 1-\frac{\langle m^4\rangle}{3\langle m^2\rangle^2}.
\end{equation}
The Binder cumulant will, like the order parameter, rise as \(\phi\) orders in the superconducting phase. At the critical coupling it is independent of system size, so curves for different system sizes will cross \cite{PhysRevLett.47.693, binder1981finite}. We use this crossing point to determine the critical coupling for the superconducting phase transition. We also measure the zero temperature equivalent of the specific heat, namely the action susceptibility given by
\begin{equation}\label{eq:action_susceptibility}
    \chi_S = \frac{1}{N}\left[\langle S^2\rangle - \langle S \rangle^2\right],
\end{equation}
where \(S\) is given by Eq.~\eqref{eq:model_charge_picture_disc}. For the superconducting transition, we expect to see a singularity in the action susceptibility at the critical coupling. 

We can measure the bulk conductivity at the superconducting phase transition, using the method presented in Refs.~\cite{PhysRevB.44.6883, PhysRevB.49.9794}. The conductivity is given by the following Kubo formula 
\begin{equation}\label{eq:Conductivity}
    \sigma_{\alpha} =\lim_{\omega_n \to 0}  \sigma_Q \frac{\rho_{\alpha}(i\omega_n)}{\omega_n},
\end{equation}
where \(\sigma_Q = (2e)^2/h\) is the conductance quantum for Cooper pairs, and the frequency dependent stiffness is given by
\begin{equation}\label{eq:freq_dep_stiffness}
    \rho_{\alpha}(i\omega_n) = \frac{1}{N}\left[\langle\epsilon_\alpha^{\phi}\rangle - \langle \tilde{I}_{\alpha}^{*}(i\omega_n)\tilde{I}_{\alpha}(i\omega_n)\rangle\right],
\end{equation}
\begin{equation}
    \epsilon_{\alpha}^\phi = J\sum_{i}\cos(\phi_{i+\alpha}-\phi_{i}),
\end{equation}
\begin{equation}
     \tilde{I}_\alpha = J\sum_{i}\sin(\phi_{i+\alpha}-\phi_{i})e^{i\tau\omega_n}.
\end{equation}
\(\tau\) is the coordinate in the imaginary time dimension, and \(\omega_n\) is the Fourier momentum along that direction. Now, in principle we could use Eq.~\eqref{eq:Conductivity} to measure the conductivity directly for any parameter regime. The limit \(\omega \to 0\) is however not accessible in our numerics, as the resolution in Fourier space depends on the number of lattice sites. We can however use finite size scaling to measure the conductivity at the superconducting phase transition. At the phase transition, the conductivity scales as \cite{PhysRevB.44.6883}
\begin{equation}\label{eq:universal_conductivity}
    \frac{\sigma(n, L/n)}{\sigma_Q} = \frac{\sigma^*}{\sigma_Q} + c\left(\frac{\alpha}{n}- \frac{n}{L}\right),
\end{equation}
where \(n\) is given by \(\omega_n = 2\pi n/L\) and \(\alpha\) is a fit parameter to account for small system sizes in our numerics. In detail, \(\alpha\) is determined by minimizing the difference between curves for \(\sigma\) of different sizes 
\begin{equation}
    \delta = \int_{-\xi}^{\xi} dx\left[\sigma_L(x)-\sigma_{L'}(x)\right]^2,
\end{equation}
where \(x=\alpha/n -n/L\) and we used a cutoff of \(\xi = 0.4\). After \(\alpha\) has been determined, the universal conductivity at the transition \(\sigma^*\) can be read off at \(\alpha/n - n/L = 0\).

\section{Results}\label{sec:Results}
In this section we present results from extensive Monte Carlo simulations on the coupled \(XY\) model presented in Eq.~\eqref{eq:model_charge_picture_disc} using the method described in Sec.~\ref{sec:Monte_Carlo_methods}. For most of the results measurements are taken over \(4\times 10^6\) Monte Carlo sweeps, up to \(16\times 10^6\) for certain parameter regimes. The system geometry is cubic in all the results (\(L=L_\tau\)). We start by presenting an overview of the SC-EM-I transition, with the quantities used to characterize these. After this we present detailed studies of the SC-EM, EM-I and SC-CSC transitions. 

\begin{figure}[htb]
    \subfloat[\label{fig:results_full_V1_range_a}]{
        \includegraphics[width=\linewidth]{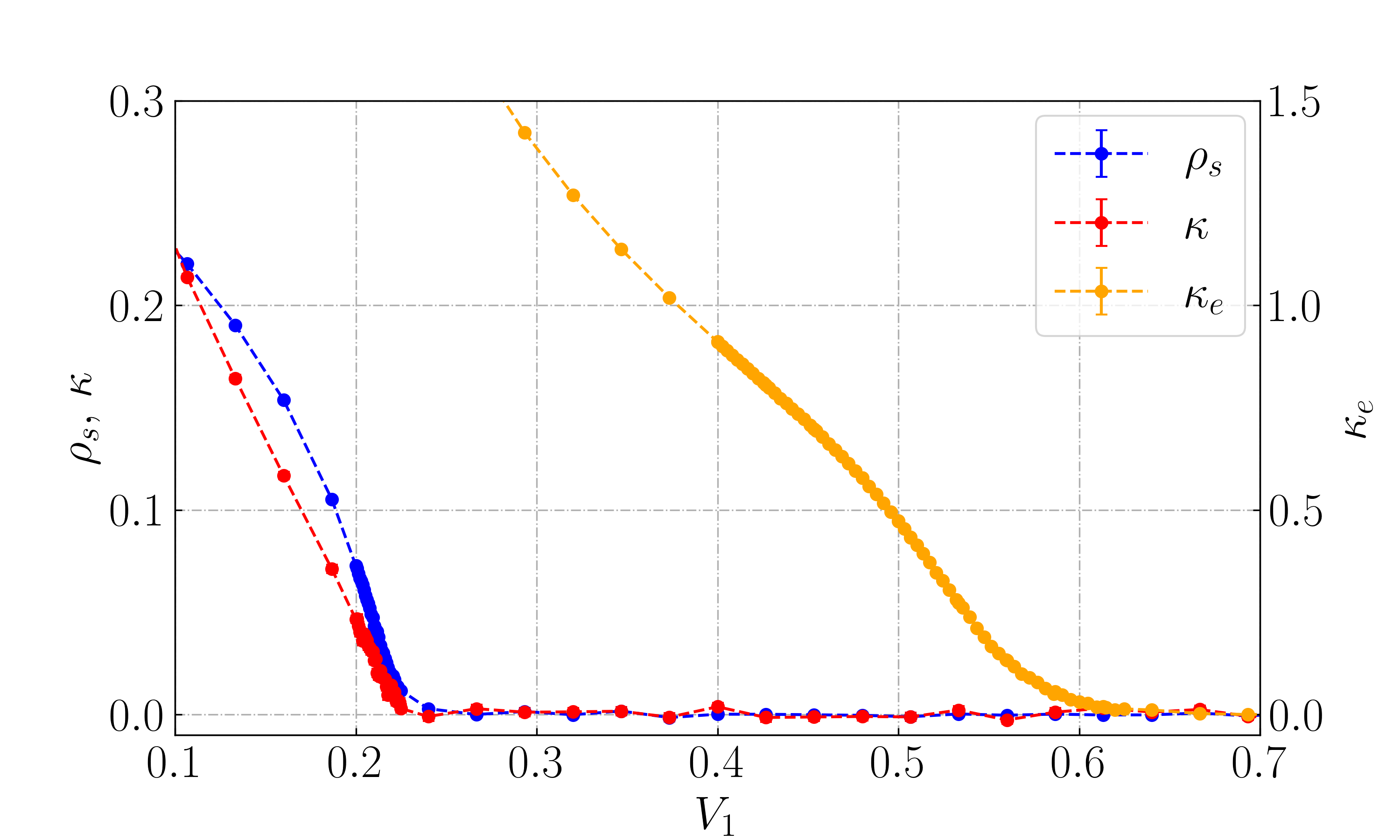}
    } \\ [-3.5ex]
    
    \subfloat[\label{fig:results_full_V1_range_b}]{
        \includegraphics[width=\linewidth]{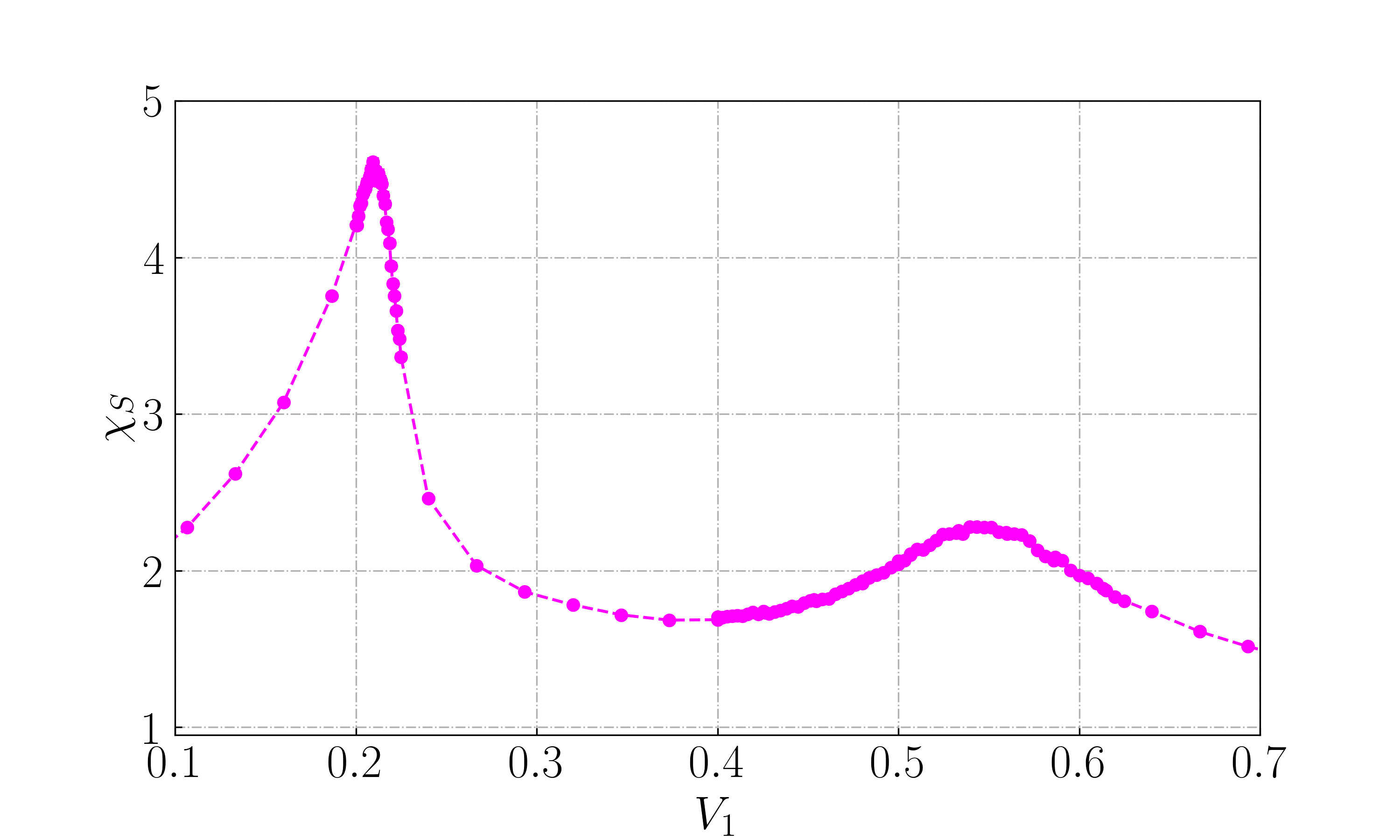}
    }
    \caption{Results from Monte Carlo simulations on the coupled \(XY\) model in Eq.~\eqref{eq:model_charge_picture_disc} with \(J=0.5\), \(V_0=0.5\) with \(L=L_\tau = 20\) (a) Superfluid stiffness given by Eq.~\eqref{eq:SF_stiffness}, bulk compressibility given by Eq.~\eqref{eq:bulk_compressibility} and edge compressibility given by Eq.~\eqref{eq:edge_compressibility}. (b) Action susceptibility given by Eq.~\eqref{eq:action_susceptibility}. Shows how the system transitions from a superconductor to an edge metal at \(V_1 \simeq 0.214\) and then from an edge metal to an insulator at \(V_1 \simeq 0.47\). Error-bars are typically smaller than the data-points.}
    \label{fig:results_full_V1_range}
\end{figure}

In Fig.~\ref{fig:results_full_V1_range} we present results for the superfluid stiffness, bulk compressibility and edge compressibility, sampled over an interval of \(V_1\) that spans the SC-EM-I transition.  For low values of \(V_1\) all three observables have a finite value. In terms of the coupled \(XY\) model, this corresponds to spatial and temporal ordering of the phase \(\phi\) and spatial ordering of the phase \(\theta\). The physical interpretation of this phase is a superconducting state, where the super-current is carried by the field \(\phi\). For intermediate \(V_1\) the superfluid stiffness and bulk compressibility drops to zero as phase coherence in \(\phi\) is destroyed by quantum fluctuations. The loss of spatial phase coherence signals the destruction of superconductivity, and the bulk of the system becomes insulating since the bulk compressibility vanishes. However, the edge compressibility remains finite which means the edges of the system can carry a current. This leads to a edge metal phase where the bulk is insulating and the edges conducting. At high values of \(V_1\), phase coherence in \(\theta\) is also lost which results in a fully insulating state. The loss of phase-coherence in $\theta$ means that the uniform susceptibility for inducing excess charge of the same sign on neighboring sites, is suppressed. This is interpreted as a state with incipient charge order. 

\subsection{Superconductor - edge metal transition}\label{subsec:SC_M_transition}
\begin{figure}[htb]
    \subfloat[\label{ig:results_PT_SC_M_a}]{
        \includegraphics[width=\linewidth]{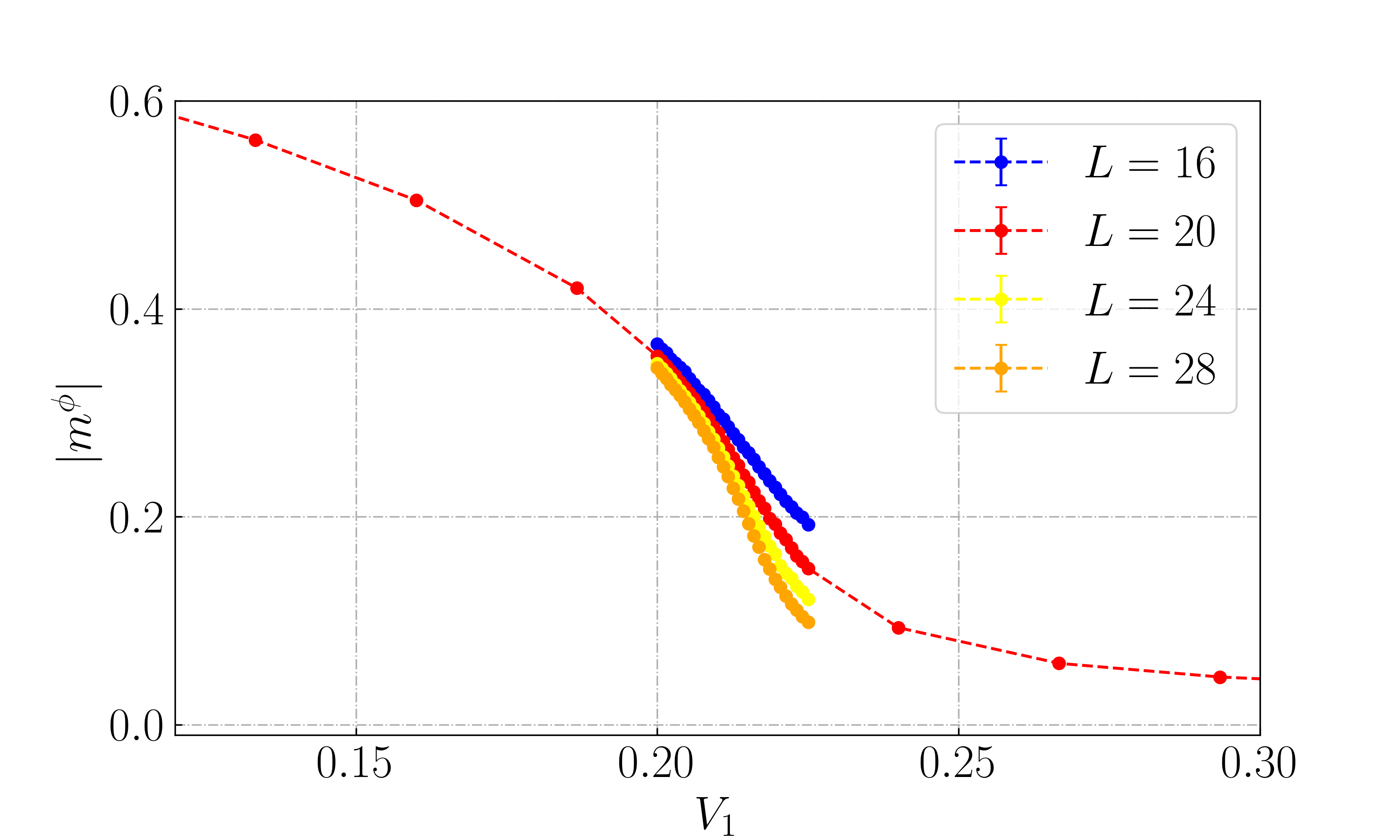}
    } \\[-3.5ex]
    
    \subfloat[\label{ig:results_PT_SC_M_b}]{
        \includegraphics[width=\linewidth]{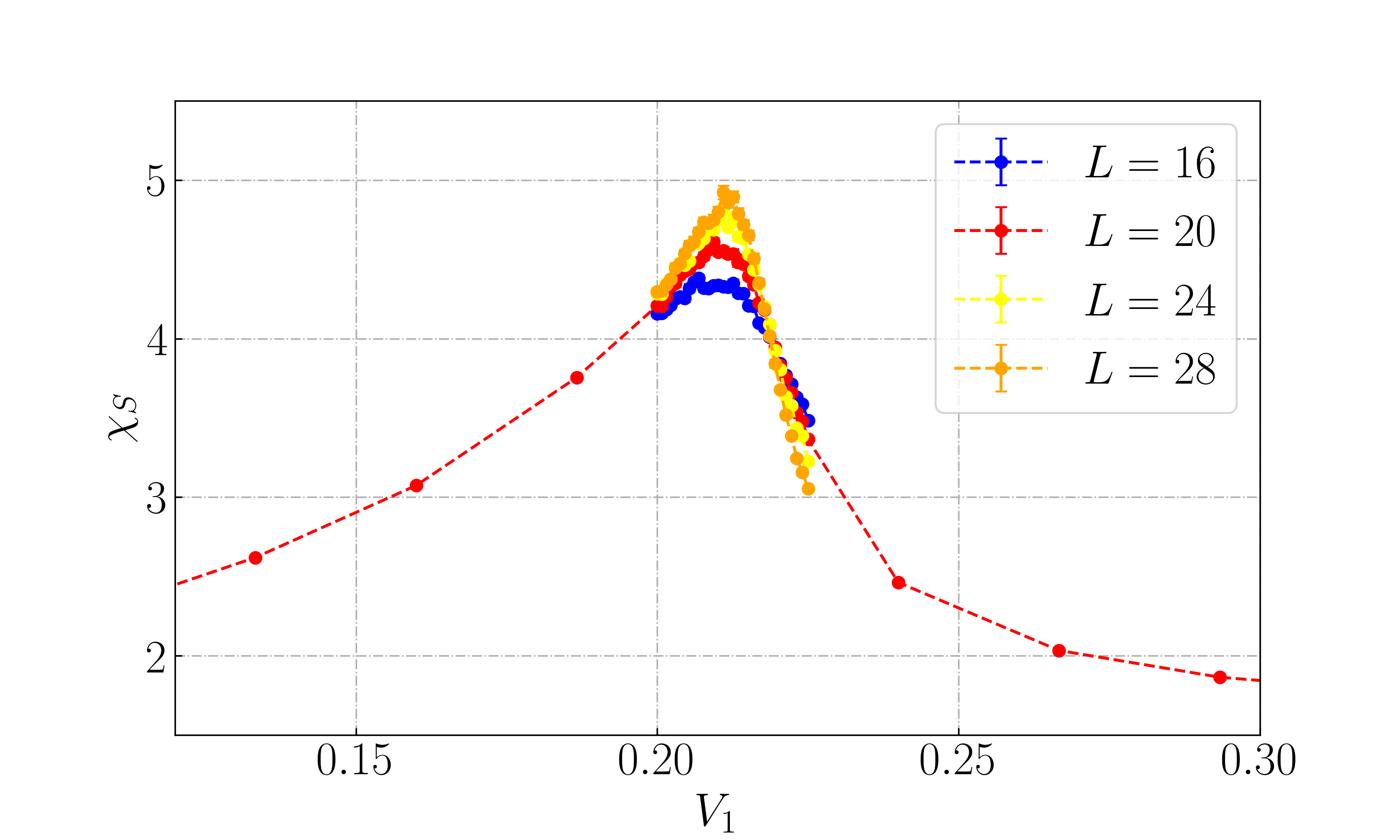}
    } \\[-3.5ex]
    \subfloat[\label{ig:results_PT_SC_M_c}]{
        \includegraphics[width=\linewidth]{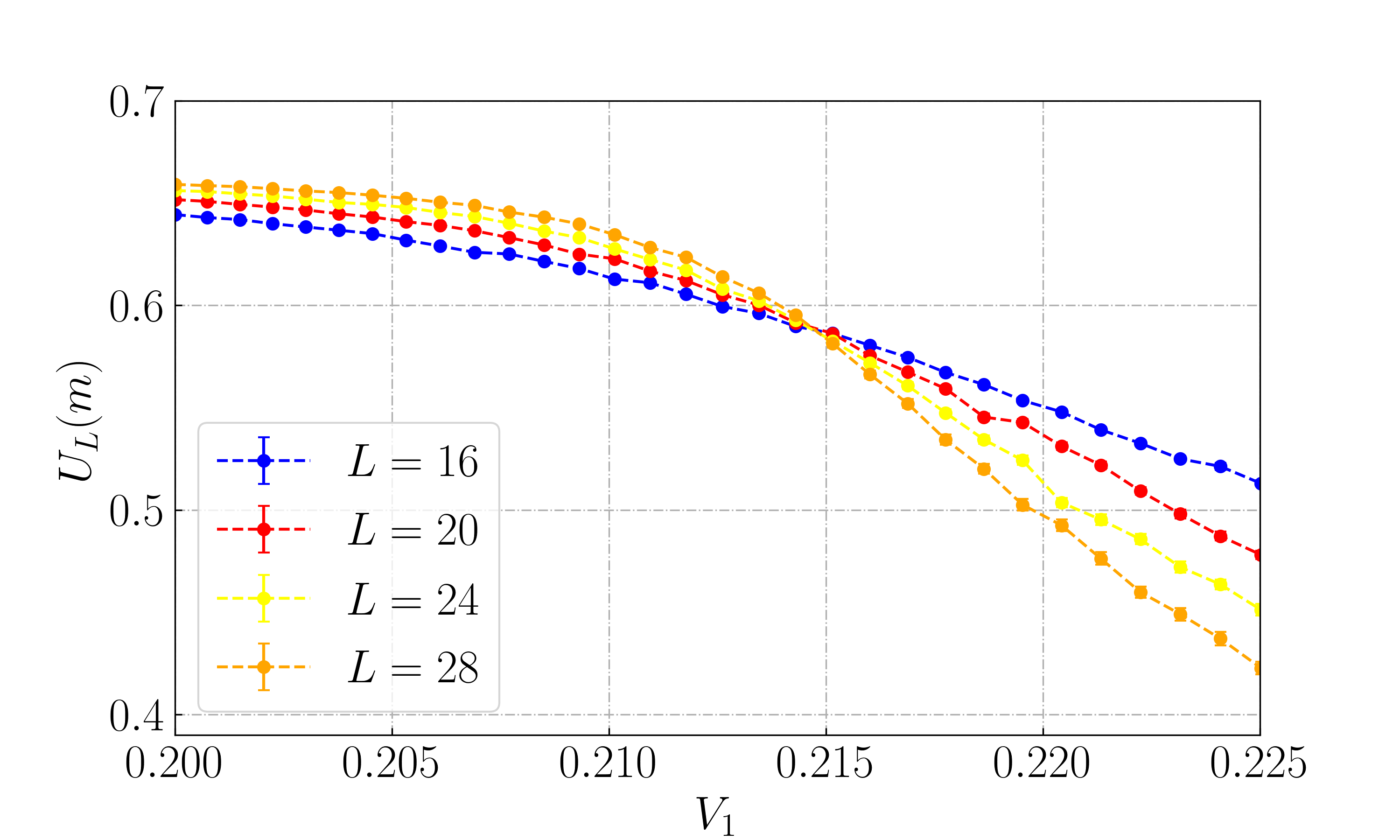}
    }
    \caption{Results for the SC-EM phase transition, using the model in Eq.~\eqref{eq:model_charge_picture_disc} with \(J=0.5\), \(V_0=0.5\) and system sizes \(L=L_\tau = 16,\, 20,\, 24,\, 28\). (a) Order parameter of \(\phi\) given by Eq.~\eqref{eq:OP_phi}. (b) Action susceptibility given by Eq.~\eqref{eq:action_susceptibility}. (c) Binder cumulant given by Eq.~\eqref{eq:Binder_cumulant}. Shows a \(3D\) \(XY\)-like phase transition at \(V_1 = 0.214\).}
    \label{fig:results_PT_SC_M}
\end{figure}

The superconductor-edge metal \ac{SC}-\ac{EM} transition is driven by spatial and temporal disordering of \(\phi\), while $\theta$ remains ordered. We can measure this through the global order parameter defined in Eq.~\eqref{eq:OP_phi}. Results for the order parameter are shown in Fig.~\ref{fig:results_PT_SC_M}, along with action susceptibility and Binder cumulant. The order parameter follows the typical behaviour we expect for a \(3D\) \(XY\) transition, where the drop becomes steeper with increasing system size. Note that the non-zero value on the metallic side of the phase transition is an artifact of our numerical method, discussed in Sec.~\ref{sec:Observables}. The drop in $|m^\phi|$ is accompanied by a peak in the action susceptibility, which becomes increasingly sharper with larger system size, typical of the \(3D\) \(XY\) phase transition. Finally, we use the Binder cumulant of the order parameter to accurately determine the critical coupling. At the phase transition, the Binder cumulant is independent of system size and we use the crossing-point of curves from different system sizes to determine the critical coupling \cite{PhysRevLett.47.693, binder1981finite}. This leads to a critical coupling \(V_1 = 0.214\) which matches the peak in the action susceptibility. The edge compressibility $\kappa_e$ remains finite above $V_1 = 0.214$, equivalently the uniform susceptibility for inducing excess charge on neighboring sites remains finite. This means that the resulting non-superconducting state remains charge-disordered.  

\begin{figure}[htb]
    \centering
    \includegraphics[width=\linewidth]{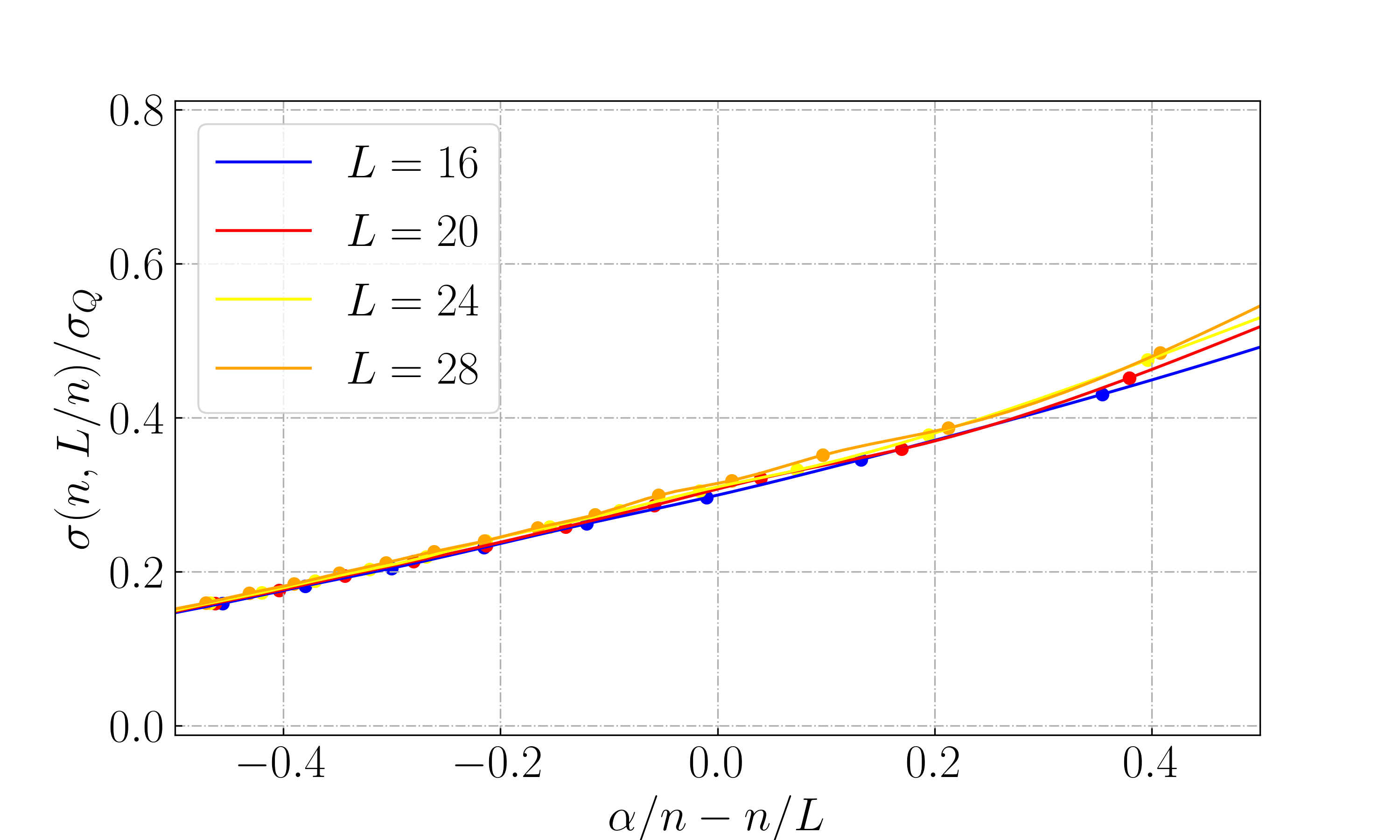}
    \caption{Conductivity (per quantum conductance) as a function of the scaling variable \(\alpha/n - n/L\), for system parameters \(J=V_0 = 0.5\) and sizes \(L=L_\tau = 16, 20, 24, 28\). The conductivity at the \ac{SC}-\ac{M} transition, \(\sigma^* / \sigma_Q\) is given by the value at \(\alpha/n - n/L = 0\).}
    \label{fig:cond_scaling}
\end{figure}

We can now use the method presented in Ref.~\cite{PhysRevB.44.6883} to calculate the bulk conductivity at the critical coupling of the SC-M phase transition. The fitting parameter in Eq.~\eqref{eq:universal_conductivity} is determined to be \(\alpha = 0.959\), from which we obtain the scaling plot in Fig.~\ref{fig:cond_scaling}. The conductivity at the transition is then given by the value at \(\alpha/n-n/L = 0\), which we determine to be
\begin{equation}
    \sigma^* = (0.31 \pm 0.01) \sigma_Q.
\end{equation}
We estimate the error by calculating the conductivity using the same method for the two data points on either side of the critical coupling on the \(V_1\)-axis. We can compare this to the value found for the large \(n\) limit of the Bose-Hubbard model in \cite{PhysRevB.44.6883}, which corresponds to the model used in this paper with \(V_1= 0\). They find \(\sigma^* = 0.285\sigma_Q\), which is argued to be universal in the sense that it only depends on the universality class of the model. In summary, we find that the \ac{SC}-\ac{EM} transition is very similar to the \ac{SC}-\ac{I} transition found in the zero temperature \(2D\) Bose-Hubbard model for large \(n\), with a \(3D\)-\(XY\) character and comparable bulk conductivity.  

\subsection{Edge metal - insulator transition}\label{subsec:EM_I_transition}
\begin{figure}[htb]
    \subfloat[\label{fig:results_PT_M_I_a}]{
        \includegraphics[width=\linewidth]{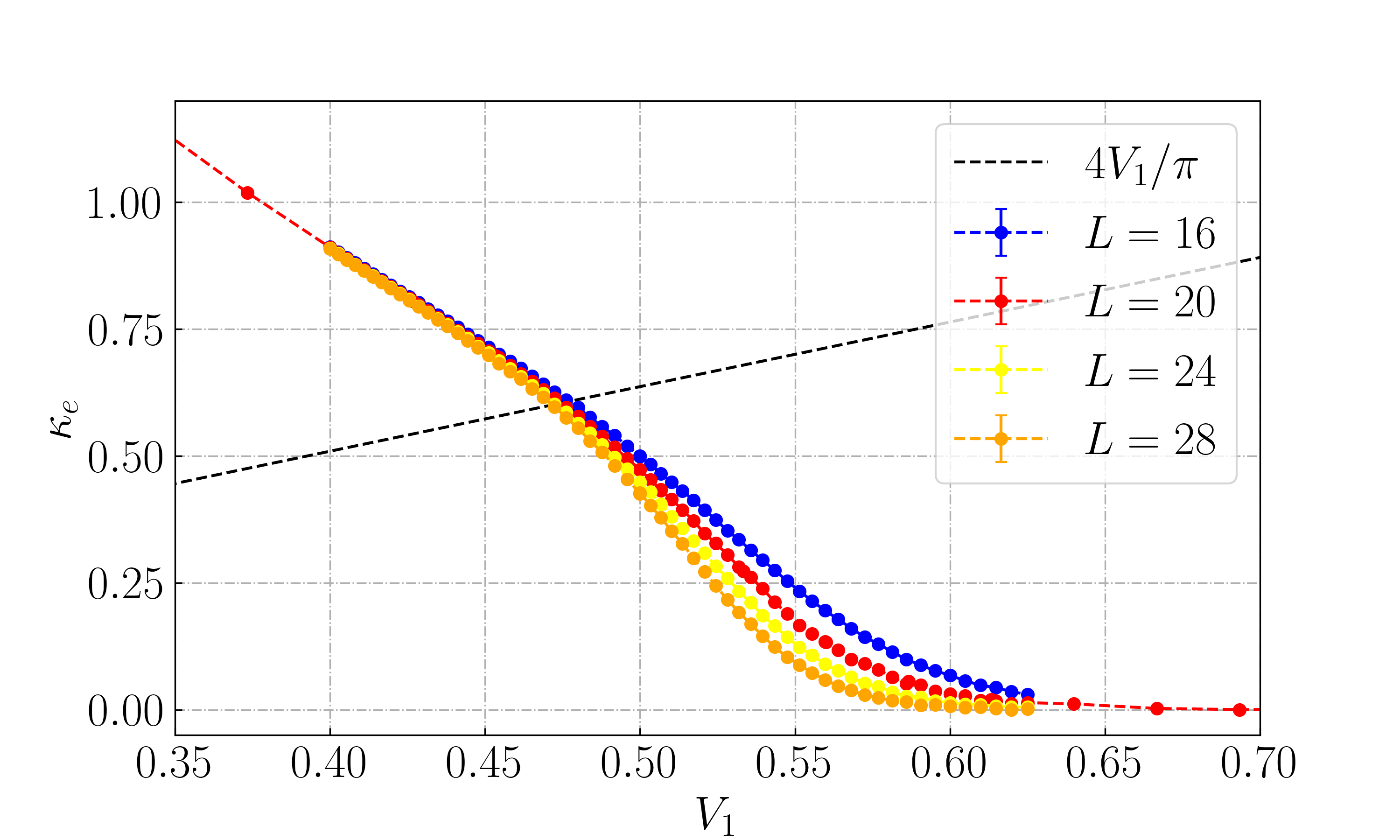}
    } \\[-3.5ex]
    
    \subfloat[\label{fig:results_PT_M_I_b}]{
        \includegraphics[width=\linewidth]{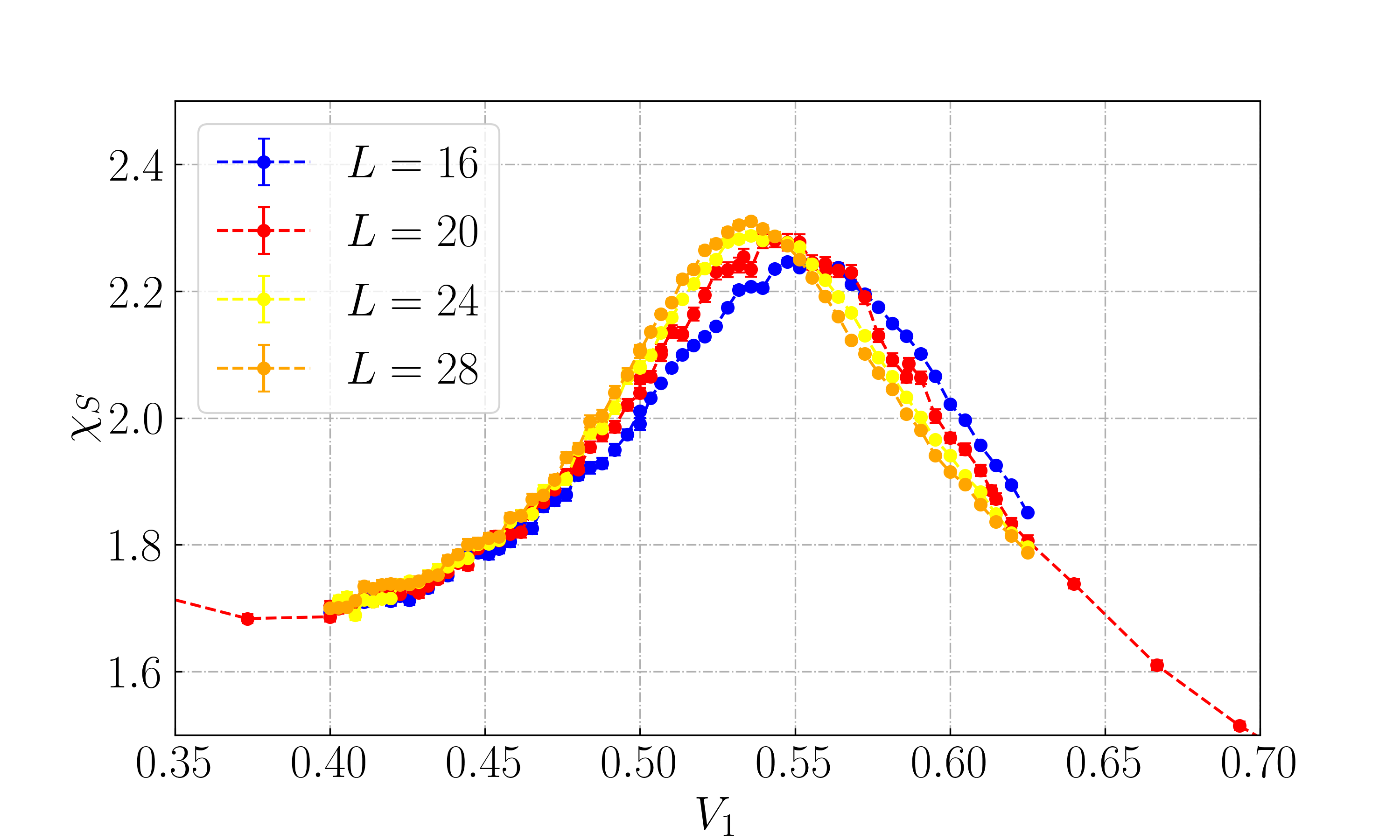}
    }
    \caption{Results for the EM-I phase transition, using the model in Eq.~\eqref{eq:model_charge_picture_disc} with \(J=0.5\), \(V_0=0.5\) and system sizes \(L=L_\tau = 16,\, 20,\, 24,\, 28\). (a) Edge compressibility given by Eq.~\eqref{eq:edge_compressibility}. (b) Action susceptibility given by Eq.~\eqref{eq:action_susceptibility}. Shows a \ac{BKT} transition at \(V_1 \simeq 0.47\).}
    \label{fig:results_PT_M_I}
\end{figure}

The edge metal-insulator \ac{EM}-\ac{I} transition is driven by spatial disordering of the phase \(\theta\). We measure this through the in-plane helicity modulus of \(\theta\),
equivalently the edge compressibility of the system. Data for the edge compressibility along with the action susceptibility at the transition is shown in Fig.~\ref{fig:results_PT_M_I}. The behaviour of both these quantities is what we would expect for the \(2D\) \(XY\) model, which is known to exhibit a \ac{BKT} phase transition \cite{kosterlitz1973ordering, berezinskii1971destruction, app11114931}. The helicity modulus is predicted to exhibit a universal jump at the transition, with magnitude \(4V_1/\pi\) in terms of the parameters used in our model. This is consistent with our data, where the helicity modulus is independent of system size at the predicted critical value of \(V_1\) (black line in Fig.~\ref{fig:results_PT_M_I}), and tends towards a jump with increasing system size. Furthermore, the action susceptibility exhibits a peak above the transition, which has also been found in previous studies of the \ac{BKT} transition \cite{app11114931}. Fig.~\ref{fig:results_full_V1_range} shows that this peak is considerably less pronounced than the one caused by the SC-EM transition. We did also consider the possibility of \(3D\)-ordering in the phase \(\theta\) through an equivalent order parameter to the one defined for \(\phi\) in Eq.~\eqref{eq:OP_phi}, which vanished as the system size was increased. Thus, our findings are consistent with a \ac{BKT} transition, and we determine the critical coupling to be the \(V_1\)-value where the predicted value of the jump matches our numerical results. For \(J=0.5\) and \(V_0 = 0.5\), the transition occurs at \(V_1 \simeq 0.47\). 

\begin{figure}[htb]
\centering
\includegraphics[width=\linewidth]{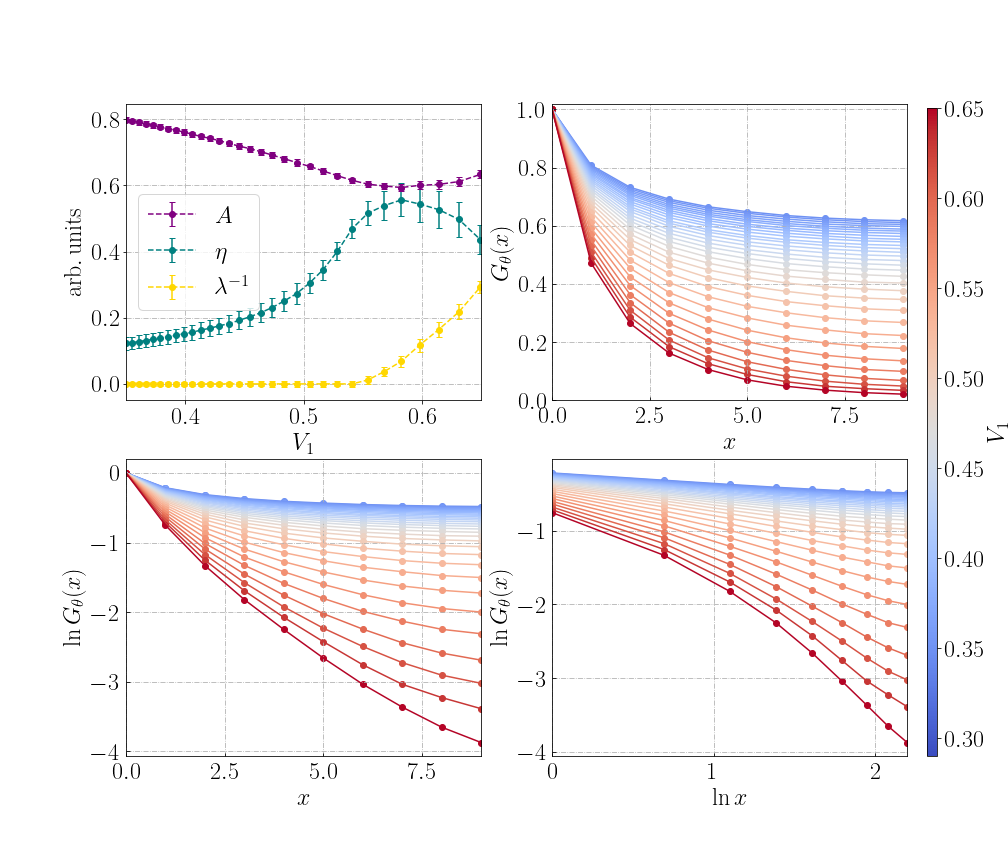}
    \caption{Measurements of the correlation function given by Eq.~\eqref{eq:correlation_function_theta}, using the model in Eq.~\eqref{eq:model_charge_picture_disc} with \(J=0.5\), \(V_0=0.5\) and \(L=L_\tau = 20\). (Top left) Three-parameter curve-fit of Monte-Carlo data to the ansatz in Eq.~\eqref{eq:CF_fit}. (Top right) Normal plot, (bottom left) log-normal plot and (bottom right) log-log plot of the correlation function for different values of \(V_1\), indicated by the color bar on the right. In all figures data has been sampled along the \(x\)-axis, with one point fixed at the origin. Data has also been averaged over all time slices (before MC average).}
    \label{fig:results_CF}
\end{figure}

One of the hallmarks of the BKT transition is the algebraic decay of the correlation function in the ordered phase, with exponential decay in the disordered phase. To investigate this, we have sampled the correlation function given in Eq.~\eqref{eq:correlation_function_theta}. Data is presented in Fig.~\ref{fig:results_CF}. The top right panel shows the correlation function for a set of \(V_1\)-values, with blue curves in the ordered phase and red curves in the disordered phase. This clearly shows that correlations grow significantly stronger across the phase transition. To investigate the possibilities of exponential and algebraic decay, we include a log-norm (bottom left) and log-log (bottom right) plot of the correlation function. The log-norm plot shows that the correlation function tends toward exponential decay for large \(V_1\) in the strongly disordered phase, and the log-log plot shows algebraic decay for small \(V_1\) in the ordered phase. Clearly, there is some cross-over between the two regions. To investigate this, we consider a general form of the correlation function given by
\begin{equation}\label{eq:CF_fit}
    G_\theta(x) = \frac{A(V_1)}{x^{\eta(V_1)}} \exp\{-x/\lambda(V_1)\}, 
\end{equation}
where in principle the amplitude \(A\), the correlation length \(\lambda\) and the coefficient \(\eta\) can all depend on \(V_1\). We fit our Monte Carlo data to this form, keeping all three parameters free. The results of this fit is shown in the upper left panel of Fig.~\ref{fig:results_CF}. Firstly we note that the correlation length diverges as we approach the phase transition from above, consistent with what we would expect for a BKT transition. Note that this occurs slightly above our estimated value of the transition, which is a finite size effect. Comparing to Fig.~\ref{fig:results_PT_M_I}, we see that the correlation length diverges around the same value that the edge compressibility drops to zero. In an infinite system, we would expect the drop to occur exactly at the transition accompanied by a diverging correlation length. The divergence of the correlation length then gives purely algebraic behaviour in the ordered phase, according to Eq.~\eqref{eq:CF_fit}. Furthermore we note that the coefficient \(\eta\) increases linearly with \(V_1\) below the transition, which is also consistent with a BKT transition. 

A limitation of the method used in this paper, is that we are not able to determine how large the edge-conductivity of the metallic state is. We note however that the conductivity is related to the compressibility through \cite{dobrosavljevic2012introduction, doi:10.1126/science.aau7063}
\begin{equation}\label{eq:edge_conductivity}
    \sigma = D\kappa,
\end{equation}
where the diffusivity \(D\) is some unknown material constant.   

\subsection{Superconductor - critical superconductor transition}\label{subsec:SC-CSC}
\begin{figure}[htb]
    \subfloat[\label{fig:results_PT_SC_SC_a}]{
        \includegraphics[width=\linewidth]{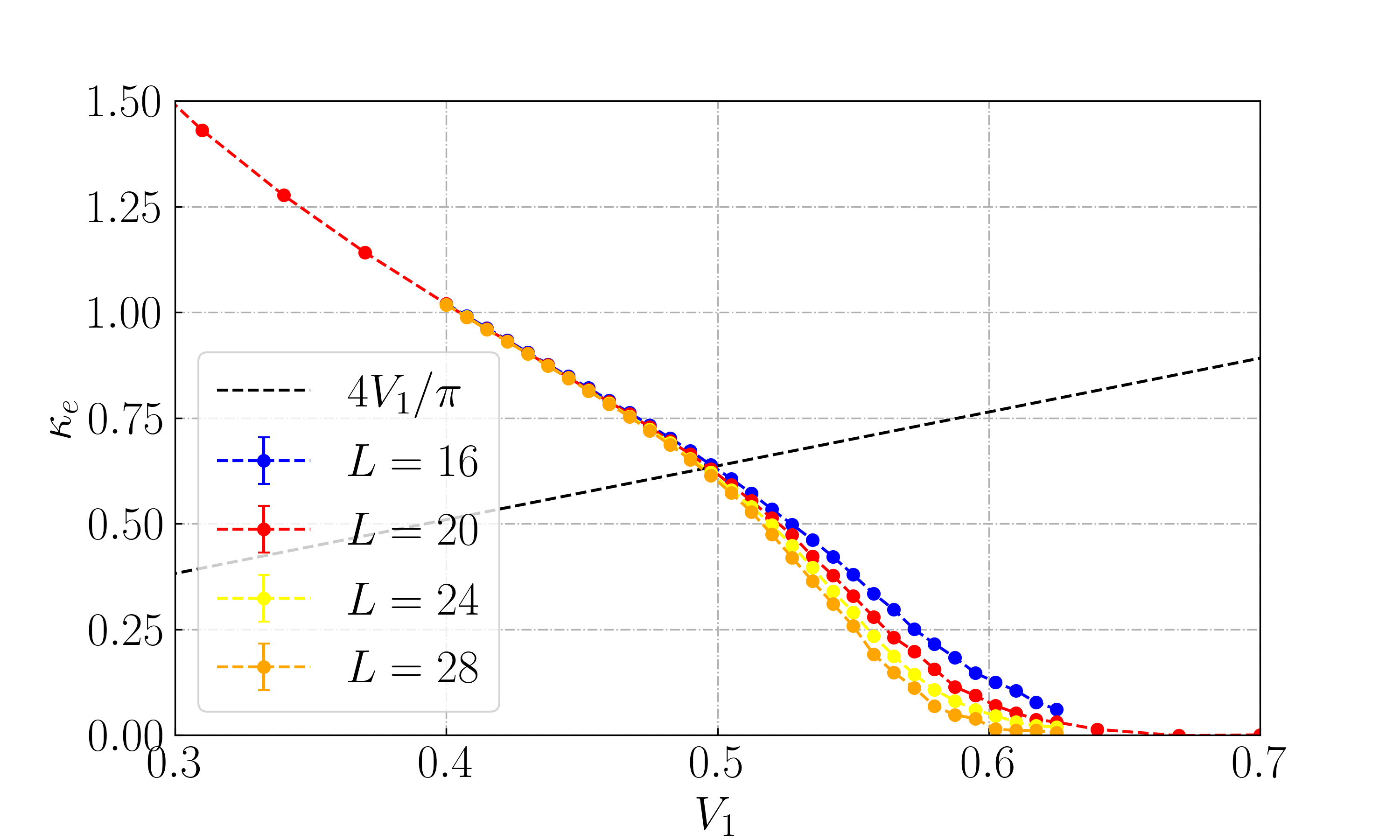}
    } \\ [-3.5ex]
    
    \subfloat[\label{fig:results_PT_SC_SC_b}]{
        \includegraphics[width=\linewidth]{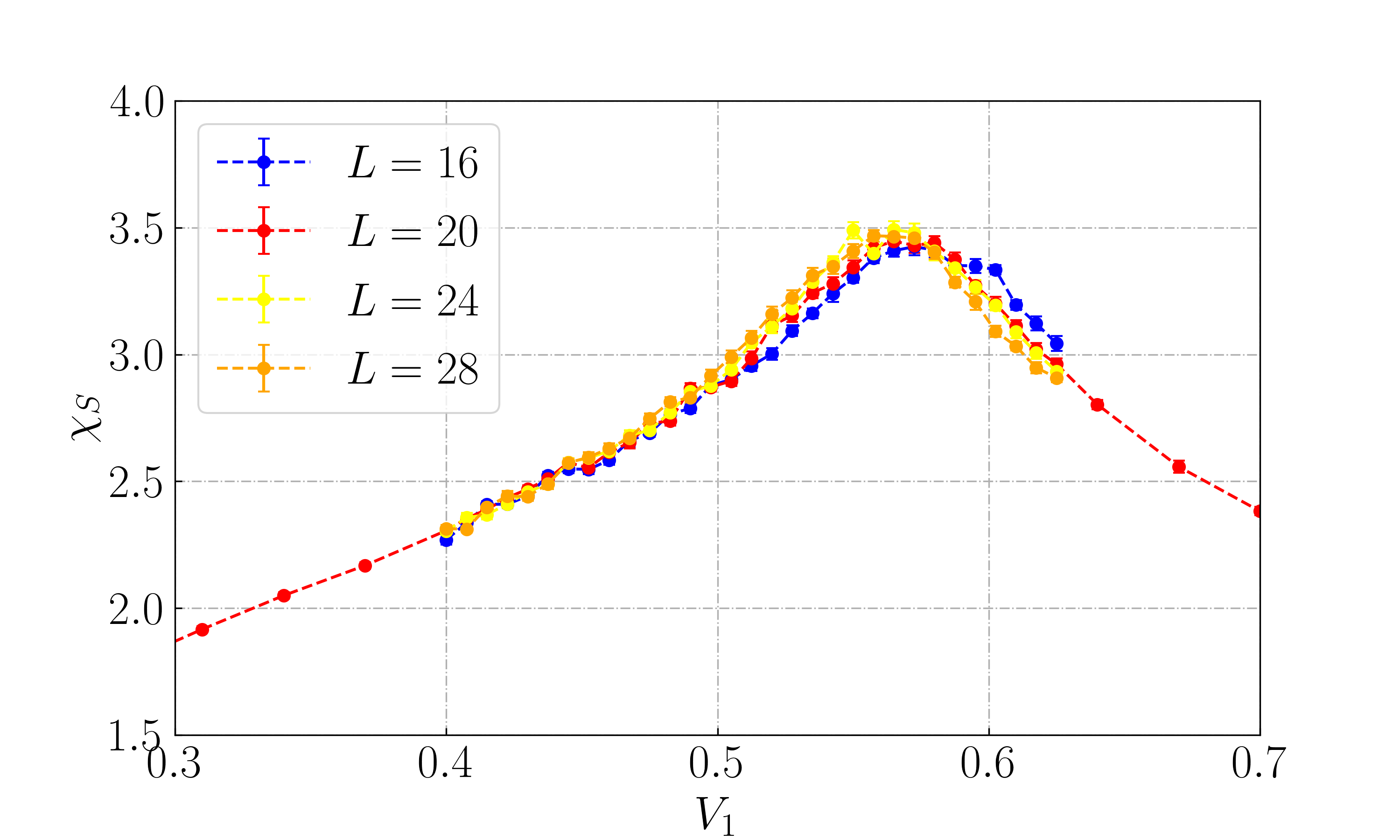}
    } \\ [-3.5ex]
    
    \subfloat[\label{fig:results_PT_SC_SC_c}]{
        \includegraphics[width = \linewidth]{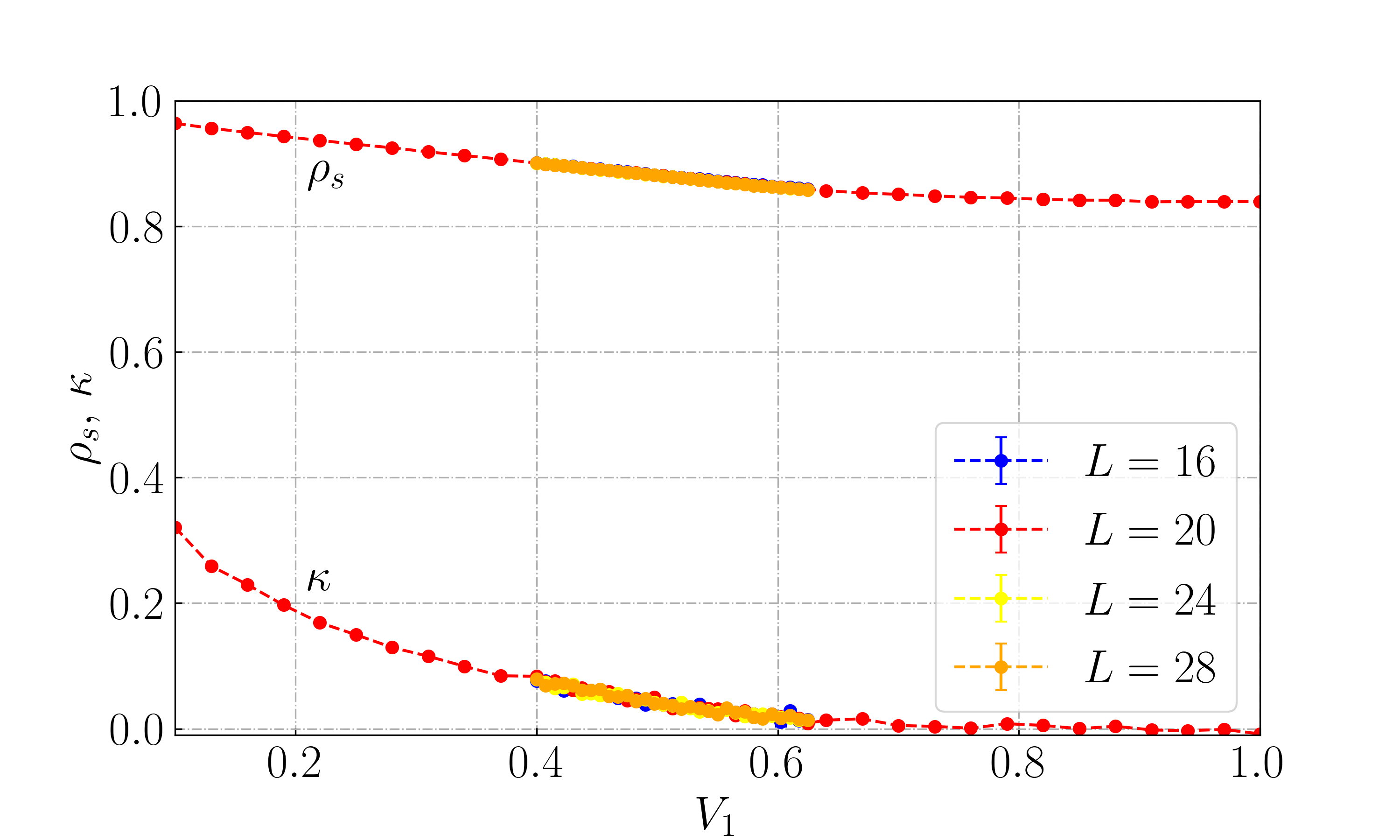}
    } \\
    \caption{Results for the SC-CSC phase transition, using the model in Eq.~\eqref{eq:model_charge_picture_disc} with \(J=1.2\), \(V_0=0.5\) and system sizes \(L=L_\tau = 16,\, 20,\, 24,\, 28\). (a) Edge compressibility given by Eq.~\eqref{eq:edge_compressibility}. (b) Action susceptibility given by Eq.~\eqref{eq:action_susceptibility}. (c) Bulk compressibility and superfluid density given by Eqs.~\eqref{eq:bulk_compressibility}, \eqref{eq:SF_stiffness}. Shows a BKT transition in \(\theta\) at \(V_1 \simeq 0.5\) along with a crossover from \(3D\) to \(2D\) ordering of \(\phi\).}
    \label{fig:results_PT_SC_SC}
\end{figure}

The superconductor-critical superconductor SC-CSC transition is driven by spatial disordering of the phase \(\theta\), similar to the to the EM-I transition discussed in Sec.~\ref{subsec:EM_I_transition}, leading to a decoupling of phases $\phi$ along the $\tau$-direction. The difference is that the SC-SCS transition occurs for a higher value of \(J > J_c \simeq 1.03\) where \(\phi\) is ordered a priori. With increasing \(V_1\) we go from the superconducting phase where \(\theta\) is \(2D\) ordered and \(\phi\) is \(3D\) ordered, to a critical superconducting phase where \(\theta\) is disordered and \(\phi\) is \(2D\) ordered in the plane but not along the temporal direction. Fig.~\ref{fig:results_PT_SC_SC} displays data for the edge compressibility and action susceptibility. These show a \ac{BKT} transition in \(\theta\) at \(V_1 \simeq 0.5\), where the edge compressibility tends towards a discontinuous jump at the critical coupling \(V_1 \simeq 0.5\) with increasing system size. Comparing with results for the EM-I transition for lower \(J\) in Fig.~\ref{fig:results_PT_M_I}, we see that both the edge compressibility and action susceptibility behave very similarly. The only notable difference is that the critical coupling has increased slightly with increasing \(J\). From this we conclude that the ordering of \(\theta\) has no qualitative dependence on whether \(\phi\) is ordered or not, at least for the parameter regime considered. 

We now turn to how the field \(\phi\) behaves at the SC-CSC transition. Results for the bulk compressibility and superfluid stiffness are shown in Fig.~\ref{fig:results_PT_SC_SC_c}. We see that the superfluid stiffness remains finite through the transition, meaning \(\phi\) is spatially ordered in both phases. The bulk compressibility is finite in the SC phase but zero in the CSC, meaning \(\phi\) disorders along the temporal direction. For the system sizes we have access to, there are no obvious signs an S-shaped curve, nor any size dependent behaviour. This leads us to conclude that the transition from \(3D\) to \(2D\) ordering of \(\phi\) is a cross-over phenomenon and not a genuine phase transition. The SC-CSC transition is still regarded a phase transition, in the sense that \(\theta\) shows a \ac{BKT} transition. To understand how \(\phi\) disorders along the temporal direction, we turn to the coupled \(XY\) model in Eq.~\eqref{eq:model_charge_picture}. As \(\theta\) disorders, we also get wild fluctuations in \(\nabla^{2}_{||}\theta\). Since this couples to the temporal gradient of \(\phi\) through the \(V_0\)-term, it forces \(\phi\) to disorder along the temporal direction while spatial order prevails due to the high value of \(J\). This is a highly unusual phenomenon, since typically in anisotropic \(XY\) models ordering in all dimensions occur simultaneously. Such a phase has been studied previously in the dissipative Josephson junction model \cite{PhysRevB.85.224531, PhysRevB.91.205129}. 

\subsection{Phase diagram overview}
In the previous sections, we have presented systematic studies of the SC-EM, EM-I and SC-SCS transitions. As the phase diagram in Fig.~\ref{fig:phase_diagram} shows, the model exhibits two more phase transitions. For intermediate values of \(J\), we find a direct SC-I transition, and for high values of \(V_1\) we find a I-CSC transition. We have not done finite size scaling of these transitions, as we already know how both fields order on either side of the transitions from the previous results. However, we have done simulations on a \(20\times20\times20\) system to determine the phase boundaries used in Fig.~\ref{fig:phase_diagram}. In this section we briefly summarize these results, which are presented in Appendix \ref{ap:More_MC_results}.

The I-CSC phase transition is driven by the spatial ordering of \(\phi\). We know that \(\phi\) is fully disordered in the insulating phase and found in Sec.~\ref{subsec:SC-CSC} that \(\phi\) is \(2D\) ordered in the critical superconducting phase. Thus, we anticipate a BKT transition in \(\phi\) where the superfluid density exhibits a universal jump with magnitude \(2/\pi J_c\) at the critical coupling \(J_c\). From the numerical results, we use the predicted value of the universal jump to determine the critical coupling, similar to what was done for the edge compressibility at the EM-I transition in Sec.~\ref{subsec:EM_I_transition}. A representative example of results for the I-CSC transition is shown in Fig.~\ref{fig:results_PT_I_CSC}.

The SC-I phase transition is driven by ordering in both phases. \(\theta\) is disordered in the insulating phase, \(2D\) ordered in the superconducting phase, and undergoes a BKT-like transition at the SC-I phase transition. \(\phi\) is also disordered in the insulating phase, \(3D\) ordered in the superconducting phase, and undergoes a \(3D\) \(XY\)-like transition. For \(J = 0.75\) we find that the two transition points coincide perfectly, for the resolution used in the simulations. The peak in specific heat associated with the \(3D\) \(XY\) phase transition matches the point where the \(\theta\) helicity modulus reaches the predicted value for the universal jump (see Fig.~\ref{fig:results_PT_SC_I}). For higher values of \(J\), as we move closer to the I-CSC transition, the ordering of \(\phi\) is harder to resolve due to strong finite size effects around the tricritical point. As shown in Fig.~\ref{fig:results_PT_I_CSC}, in a finite system the superfluid density will remain finite in the insulating phase close to the critical superconducting phase. As we move into the insulating phase from the superconducting phase, close to the critical superconducting phase, the superfluid density will not go to zero but remain finite. This is however not an issue for the \(\theta\)-field, which shows a clear phase transition for all values of \(J\) in the phase diagram in Fig.~\ref{fig:phase_diagram}. Consequently, we argue that in the thermodynamic limit the SC-I phase transition should be a simultaneous \(3D\) \(XY\) in \(\phi\) and BKT in \(\theta\) phase transition for \(J \in (0.75, 1.0)\), and use the critical coupling of the BKT transition in \(\theta\) to determine the phase boundary.

\section{Summary and discussion}\label{sec:Summary_discussion}
In this paper we have investigated a quantum rotor model representation of the extended Hubbard model in the case where there are many bosons per site. This has been proposed as a candidate theory to describe the \ac{SC}-\ac{M}-\ac{I} transition found at zero temperature in two-dimensional systems. We have shown that the model exhibits four distinct phases. Namely, i) a superconducting phase characterized by phase coherence in the field \(\phi\), ii) an edge metal phase which is insulating in the bulk with conducting edge states characterized by spatial phase coherence in the field \(\theta\), iii) a critical superconducting phase characterized by spatial phase coherence in \(\phi\) and temporal disordering in \(\phi\), and iv) an insulating phase characterized by disorder in both fields. We start by comparing our results to the findings in Ref.~\cite{PhysRevB.60.1261} on the same model. One of the major differences is the transition line where \(\theta\) disorders. We find that it is essentially independent of both \(V_0\) and \(J\) for the parameters considered, which leads to a very different phase diagram. We can understand this from the coupled \(XY\) model, where the gradient term plays the dominant role in ordering \(\theta\). In other words, even when fluctuations in \(\phi\) along the temporal direction are completely suppressed, the \(V_0\) term cannot order \(\theta\) by itself. An equivalent statement is that ordering of \(\phi\) has very little impact on the ordering of \(\theta\). The inverse statement is not true, as both phase diagrams in Fig.~\ref{fig:phase_diagram} shows that ordering in \(\phi\) can be achieved by tuning only \(V_1\). To understand this consider the limit \(V_1 \to 0\) where fluctuations in \(\theta\) are completely frozen out and the coupled \(XY\) model reduces to the following anisotropic \(3D\) \(XY\) model for the phase \(\phi\)
\begin{equation}
    S = -J\sum_{i,\alpha}\cos(\nabla_\alpha \phi_i) - \frac{1}{2V_0}\sum_i \cos(\nabla_\tau\phi_i).
\end{equation}
This features long range order in \(\phi\) for sufficiently small \(V_0/J\), and in Fig.~\ref{fig:phase_diagram} the transition occurs for \(V_0 = 1.32\), \(J=0.5\). Increasing \(V_1\) will increase fluctuations in \(\theta\), acting to disorder \(\phi\) due to the \(V_0\) coupling term. At some intermediate value of \(V_1\) the fluctuations in the \(\nabla^2_{||}\theta\) term will be sufficiently strong to disorder \(\phi\) completely by effectively renormalizing the coupling constant along the temporal direction, resulting in a decoupling transition in the $\tau$-direction. Such a decoupling transition cannot happen in a single-component anisotropic $XY$-model. Since $J=0.5 < J_{c2D}$, this results in complete disordering of $\phi$. Here, $J_{c2D}$ is the critical value of $J$ below which the $2DXY$ model disorders.  As \(V_0 /J\) is reduced larger fluctuations in \(\nabla^2_{||}\theta\) are required to disorder \(\phi\), which is consistent our numerical results. 

Our results also lead to a different understanding of the intermediate phase than Ref.~\cite{PhysRevB.60.1261}. We find that quantum fluctuations in \(\phi\) result in an incompressible bulk in the intermediate phase, meaning charge fluctuations in the bulk excitation spectrum are gapped. The only surviving observable in the intermediate phase is the non-zero edge compressibility. Thus, although the bulk is insulating the edges can conduct a resistive current. A deficit of our method is that we are not able to estimate the edge conductance, which makes it hard to compare our results directly to experiments. Thus, this paper should be viewed as a proof of principle of how an edge metal state can be realized in two dimensions, starting from a relatively simple description of superconducting grains with basic interactions. We can also provide some qualitative insight into how the metallic and superconducting degrees of freedom coexist. As shown in Sec.~\ref{subsec:SC_M_transition}, the \ac{SC}-\ac{EM} transition is very reminiscent of the \ac{SC}-\ac{I} transition described by the Bose-Hubbard model without the nearest neighbor \(V_1\)-term \cite{PhysRevB.44.6883}. This leads us to hypothesize that the superconducting and metallic degrees of freedom can be viewed as two conducting channels (bulk and edge) in parallel coupling, so that the overall resistance of the system is given by
\begin{equation}
    R = \frac{R_b R_e}{R_b + R_e}.
\end{equation}
In the \ac{EM} phase \(R_b \to \infty\) so that the overall resistance is \(R = R_e\). As we approach the insulating phase the edge resistance will increase according to Eq.~\eqref{eq:edge_conductivity}, which can qualitatively explain the smooth evolution of the sheet resistance across the intermediate metallic regime found in experiments. 

We also found that the model exhibits a fourth phase, the CSC phase where \(\theta\) is disordered and \(\phi\) is \(2D\) ordered. This phase is also hypothesized in Ref.~\cite{PhysRevB.60.1261} as a possible supersolid phase. We do not find any stable charge density wave ordering, since fluctuations in \(m\) are always suppressed when \(\phi\) is disordered. However we showed in Sec.~\ref{sec:Observables} that disordering \(\theta\) results in incipient charge density wave ordering. Therefore, this phase is closely related to a supersolid, with a finite superfluid density, but the charge ordering is never realized due to suppression of all charge fluctuations. Instead we identify this phase as a critical superconducting phase, which has been realized earlier in resistive Josephson junction arrays \cite{PhysRevB.85.224531, PhysRevB.91.205129}. 

Finally, we would like to draw attention to some experimental papers on \ch{MoS2}. In Ref.~\cite{PhysRevLett.87.196803}, scanning tunneling microscopy was used to investigate single layer nanoclustered \ch{MoS2}. They reported a brim of very high conductance extending all the way around the edges of the clusters. These results were corroborated by DFT-calculations were up to two intrinsic metallic edge states were found, associated with subtle changes in the electronic structure close to the edge. A subsequent study \cite{yang2020possible} demonstrated that the edge transport plays a dominant role over bulk transport at low temperatures down to 6.3K.  Other work \cite{doi:10.1126/science.1228006} investigated the conducting properties of atomically thin layers of \ch{MoS2} tuned by gate voltages. They reported an intermediate metallic phase between the superconducting and insulating regime  at very low temperatures. Whether the metallic edge states found in Ref.~\cite{PhysRevLett.87.196803} can explain the low temperature metallic behaviour found in Ref.~\cite{doi:10.1126/science.1228006}, and more generally whether the mechanism we demonstrate in this paper can explain all the experimental reports of zero temperature metallic behaviour, still remains very much an open question. However, we believe that our results call for a systematic study of the local conducting properties for a wider parameter regime spanning all three phases of these systems to illuminate the issue further. 

\section{Acknowledgements}
We acknowledge financial support from the Research Council of Norway Grant No. 262633 “Center of Excellence on Quantum
Spintronics,” and Grant No. 323766. We thank C. M. Varma for several helpful discussions.

\bibliography{references.bib}
\appendix
\begin{widetext}
\section{Bulk compressibility}\label{ap:Bulk_compressibility}
In this appendix, we show that the bulk compressibility is equal to the helicity modulus of \(\phi\) along the temporal direction. This will also partly include the derivation of the coupled \(XY\) model in Eq.~\eqref{eq:model_charge_picture} starting from Eq.~\eqref{eq:JJA_Hamiltonian}, but see Ref.~\cite{PhysRevB.60.1261} for more details. Starting from Eq.~\eqref{eq:JJA_Hamiltonian}, we write down the partition function in the coherent path integral formalism.
\begin{equation}
    Z = \sum_{\{m_i\}} \int_0^{2\pi}\mathcal{D}\phi e^{-S},
\end{equation}
\begin{equation}\label{eq:action_with_chem_pot}
    S = i\sum_{i}m_i(\nabla_\tau\phi_i) - J\sum_{i\alpha} \cos(\phi_i -\phi_{i+\alpha}) + V_0\sum_{i}m_i^2 + V_1\sum_{i\alpha}(m_i + m_{i+\alpha})^2 - \mu\sum_i m_i,
\end{equation}
where we have set \(\Delta\tau = 1\). The expectation value of the occupation number is now given by
\begin{equation}
    \langle m \rangle \equiv \frac{1}{Z} \sum_{\{m_i\}}\int_0^{2\pi} \mathcal{D}\phi \left(\frac{1}{N}\sum_i m_i\right)e^{-S} = \frac{1}{N}\frac{1}{Z}\frac{\partial Z}{\partial\mu}.
    \label{average m}
\end{equation}
which is zero if $\mu =0$, but could differ from zero if $\mu \neq 0$. This implies immediately that $\partial \langle m \rangle/\partial \mu$ could differ from zero even if $\mu =0$.   
Inserting Eq. \ref{average m} into the definition of the compressibility  Eq.~\eqref{eq:bulk_compressibility}, we find
\begin{equation}\label{eq:compressibility_S}
    \kappa = \frac{1}{N}\left[-\left\langle \frac{\partial S}{\partial \mu}\right\rangle^2 - \left\langle \frac{\partial^2 S}{\partial \mu^2}\right\rangle + \left\langle\left(\frac{\partial S}{\partial \mu}\right)^2 \right\rangle\right]\Bigg\vert_{\mu = 0}.
\end{equation}
The first term is current-like and will have zero expectation value in both the ordered and disordered phases, so we neglect it in the main text result. To measure the compressibility using the coupled \(XY\) model in Eq \eqref{eq:model_charge_picture_disc}, we then need to carry out the calculation from Eq.~\eqref{eq:action_with_chem_pot}. By rotating the integer field \(m_i \to m_i e^{iQ\cdot r_i}\), where \(Q = (\pi, \pi)\), the action becomes
\begin{alignat}{1}
S =& i\sum_{i}e^{-iQ\cdot r_i}m_i(\nabla_\tau \phi_i) - J\sum_{i\alpha} \cos(\phi_i -\phi_{i+\alpha}) + V_0\sum_i m_i^2 + V_1 \sum_{i\alpha}(\nabla_\alpha m_i)^2 - \mu\sum_i e^{-iQ\cdot r_i}m_i. 
\end{alignat}
Note how the $V_1$-term has been turned into a gradient of the rotated $m_i$.  
The \(V_1\) term can be rewritten using a Hubbard Stratonovich transformation
\begin{alignat}{1}
    \exp\left[-V_1\sum_{i\alpha}(\nabla_\alpha m_i)^2\right] \sim & \int_{-\infty}^{\infty}\mathcal{D}p \exp\left[-\frac{1}{4V_1}\sum_{i\alpha}p_{i,\alpha}^2 - i \sum_{i}m_i(\nabla_\alpha p_{i,\alpha})\right] \label{eq:HS_bulk_comp} \\
   = & \sum_{l_{i,\alpha}}\int_{0}^{2\pi}\mathcal{D}\theta \exp\left[-\frac{1}{4V_1}\sum_{i\alpha}(\nabla_\alpha \theta_i -2\pi l_{i,\alpha})^2 - i \sum_{i}m_i(\nabla^2 \theta_i)\right] \nonumber\\
   \simeq & \int_{0}^{2\pi}\mathcal{D}\theta \exp \left[\frac{1}{2V_1}\sum_{i\alpha}\cos(\nabla_\alpha \theta_i)- i \sum_{i}m_i(\nabla^2 \theta_i)\right]. \nonumber
\end{alignat}
The intermediate steps involve splitting the integral over \(p_i\) into intervals of \(2\pi\), by introducing a phase and an integer field. This phase can be written as the sum of a gradient and a curl, where the curl vanishes from the dynamics since only the gradient of \(p\) couples to \(m\). The final step inverts the Villain approximation. This part of the derivation is explicitly included to show how the field \(\theta\) comes from decoupling the densities at neighbouring sites. The action now reads 
\begin{alignat}{1}
    S = &-\sum_{i\alpha}\cos(\phi_i - \phi_{i+\alpha}) -\frac{1}{2V_1}\sum_{i\alpha}\cos(\theta_i - \theta_{i+\alpha})\\
    &+ i \sum_{i}m_i(e^{-iQ\cdot r_i}(\nabla_\tau \phi_i) + \nabla^2\theta_i + i  e^{-iQ\cdot r_i}\mu) + V_0\sum_i m_i^2, \nonumber
\end{alignat}
We can now integrate out the field \(m\) by completing the square. Using again the inverse Villain approximation, we obtain the action in terms of two coupled \(XY\) models
\begin{equation}
    Z = \int_0^{2\pi}\mathcal{D}\phi   \mathcal{D} \theta e^{-S},
\end{equation}
\begin{equation} \label{eq:action_charged_mu}
     S = -J\sum_{i\alpha}\cos(\phi_i - \phi_{i+\alpha}) -\frac{1}{2V_1}\sum_{i\alpha}\cos(\theta_i - \theta_{i+\alpha}) - \frac{1}{2V_0}\sum_i \cos(e^{-iQ\cdot r_i}(\nabla_\tau \phi_i) + \nabla^2\theta_i + i  e^{-iQ\cdot r_i}\mu).
\end{equation}
We see that we obtain Eq.~\eqref{eq:model_charge_picture} in the main text by setting \(\mu = 0\). The final Villain approximation here is not entirely trivial, since the argument of the cosine is complex. However, since \(\mu\) is fixed, the complex part is just a shift along the imaginary axis and the approximation is still valid. Inserting the action in Eq.~\eqref{eq:action_charged_mu} into the expression for the compressibility in Eq.~\eqref{eq:compressibility_S}, we obtain the two terms for the helicity modulus in Eqs.~\eqref{eq:epsilon_phi_tau} and \eqref{eq:I_phi_tau}.

\section{Edge compressibility}\label{ap:Edge_compressibility}
In this section we show that the edge compressibility is equal to the helicity modulus of \(\theta\) along the in-plane directions. Our starting point is the action in Eq.~\eqref{eq:action_with_chem_pot} (with \(\Delta\tau = 1\))
\begin{equation}
    Z = \sum_{\{m_i\}} \int_0^{2\pi}\mathcal{D}\phi_i e^{-S},
\end{equation}
\begin{equation}\label{eq:action_with_chem_pot_boundary}
    S = i\sum_{i}m_i(\nabla_\tau\phi_i) - J\sum_{i\alpha} \cos(\phi_i -\phi_{i+\alpha}) + V_0\sum_{i}m_i^2 + V_1 \sum_{i\alpha}(m_i + m_{i+\alpha})^2 - \mu_{e, \beta} M_{\beta} - \mu_{b}\sum_{i}{}^{''}m_i,
\end{equation}
where we have split the chemical potential term into a bulk contribution denoted by the double primed sum, and the edge contribution which is defined in Eq.~\eqref{eq:edge_occupation_number}. We set \(\mu_b\) to zero, and consider an infinitesimal \(\mu_{e,\beta}\) which will also be set to zero at the end of the derivation. The expectation value of the edge occupation number is now given by
\begin{equation}
    \langle M_\alpha \rangle = \frac{1}{N}\frac{1}{Z}\frac{\partial Z}{\partial \mu_{e,\alpha}},
\end{equation}
and inserting this into the definition of the edge compressibility in Eq.~\eqref{eq:edge_compressibility} gives
\begin{equation}\label{eq:edge_compressibility_S}
    \kappa_{e,\alpha} = \frac{1}{N}\left[-\left\langle \frac{\partial S}{\partial \mu_{e,\beta}}\right\rangle^2 - \left\langle \frac{\partial^2 S}{\partial \mu_{e,\beta}^2}\right\rangle + \left\langle\left(\frac{\partial S}{\partial \mu_{e,\beta}}\right)^2 \right\rangle\right]\Bigg\vert_{\mu_{e,\beta} = 0}.
\end{equation}
The objective is now to follow similar steps as in Appendix~\ref{ap:Bulk_compressibility} to see where \(\mu_{e,\beta}\) will appear in the coupled \(XY\) model. We start by making the variable shift
\begin{equation}\label{eq:mu_shift}
    \mu_{e,\beta} \to i\gamma_\beta.
\end{equation}
Next, we rewrite the edge contribution of the chemical potential in the following manner
\begin{equation}\label{eq:M_transformed}
    M_{\beta} = \sum_{i}{}^{'} (m_{1,i} + m_{L,i}) = \sum_{i}^{L-1}{}^{'}(m_i + m_{i+\beta})e^{iQ\cdot r_i} 
\end{equation}
with \(Q= (\pi,\pi)\). The sum in the right-most expression now runs over all lattice sites except those at one edge. Carrying out the sum, the alternating phase factor will cancel any bulk contribution leaving us with exactly \(M_\alpha\). Note that the relative sign of \(m_{1,i}\) and \(m_{L,i}\) seemingly depends on whether \(L\) is even (+) or odd (-). However, the transformation of \(m\) with the phase factor \(\exp(-iQ\cdot r_i)\) is only sensible for an even system, as it introduces frustration effects in an odd system with periodic boundary conditions. The next step is reintroducing periodic boundary conditions, by extending the sum in Eq.~\eqref{eq:M_transformed} to include the edge layer. This might seem arbitrary as the sum should now be zero, but another transformation will show that it still corresponds to a boundary term. This extension to periodic boundary conditions, where the two edges are connected, is also exactly what is done when measuring the helicity modulus of the \(XY\) model. We conceptualize and explain it in terms of a finite system, where there is an infinitesimal twist introduced between every site, but in numerical simulations with periodic boundary conditions the twist is also applied to the links connecting the two edges. This point also highlights the fact that the edge compressibility is a global observable, like the helicity modulus. After reintroducing periodic boundary conditions and making the shift in Eq.~\eqref{eq:mu_shift}, the action now reads
\begin{equation}\label{eq:action_m_with_gamma}
    S = i\sum_{i}m_i(\nabla_\tau \phi_i) -J\sum_{i\alpha} \cos(\phi_i - \phi_{i+\alpha}) + V_0 \sum_i m_i^2 + V_1\sum_{i\alpha}(m_i + m_{i+\alpha})^2 
    + i\sum_i (m_i + m_{i+\beta})e^{iQ\cdot r_i}\gamma_\beta.
\end{equation}
Next, we rotate the integer field \(m_i \to m_i \exp(-iQ\cdot r_i)\) and obtain 
\begin{equation}\label{eq:action_m_with_gamma_rotated}
    S =  i\sum_{i}e^{-iQ\cdot r_i}m_i(\nabla_\tau \phi_i) - J\sum_{i\alpha} \cos(\phi_i -\phi_{i+\alpha}) + V_0\sum_i m_i^2 + V_1 \sum_{i\alpha}(\nabla_\alpha m_i)^2 - i\sum_{i\alpha} (\nabla_\alpha m_i)\gamma_\alpha',
\end{equation}
where we have introduced \(\gamma_\alpha' = \gamma_\beta \delta_{\alpha\beta}\). Note that integration by parts will show that the last term is indeed a boundary term. We can now Hubbard-Stratonovic decouple the two terms including \(\nabla_\alpha m_i\) similarly to what is done in Eq~\eqref{eq:HS_bulk_comp}
\begin{alignat}{1}
    \exp\left[- V_1 \sum_{i\alpha}(\nabla_\alpha m_i)^2 + i\sum_{i\alpha} (\nabla_\alpha m_i)\gamma_\alpha'\right] & = \int_{-\infty}^{\infty} \mathcal{D}p \exp \left[-\frac{1}{4V_1}\sum_{i\alpha}(p_{i\alpha} + \gamma'_\alpha)^2 + i \sum_{i\alpha} p_{i\alpha} \nabla_\alpha m_i\right] \\
        &= \sum_{l_{i\alpha}}\int_0^{2\pi}\mathcal{D}\theta \exp  \left[-\frac{1}{4V_1}\sum_{i\alpha}(\nabla_\alpha\theta_i - 2\pi l_{i\alpha}  + \gamma'_\alpha)^2 + i \sum_{i\alpha} \nabla_\alpha m_i \nabla_\alpha \theta_i \right] \\
        &\simeq \int_0^{2\pi}\mathcal{D}\theta \exp\left[ \frac{1}{2V_1} \sum_{i\alpha} \cos(\nabla_\alpha \theta_i + \gamma_\alpha') +i \sum_{i,\alpha} \nabla_\alpha m_i \nabla_\alpha \theta_i \right].
\end{alignat}
The difference is that we complete the square before decoupling, which leads to a shift in the squared term after decoupling and ultimately to a shift in the cosine term for \(\theta\). The final steps are identical to Appendix~\ref{ap:Bulk_compressibility}, and we thus obtain the action of two coupled \(XY\) models
\begin{equation}
    Z = \int_0^{2\pi}\mathcal{D}\phi   \mathcal{D} \theta e^{-S}.
\end{equation}
\begin{equation}\label{eq:action_charged_gamma}
    S = -J\sum_{i\alpha} \cos(\phi_i - \phi_{i+\alpha}) - \frac{1}{2V_1}\sum_{i\alpha}\cos(\theta_i - \theta_{i+\alpha} + \gamma_\alpha') - \frac{1}{2V_0}\sum_{i}\cos(e^{-iQ\cdot r_i}(\nabla_\tau \phi_i) + \nabla^2 \theta_i).
\end{equation}
Notice that \(\gamma_\alpha'\) enters the action exactly as a twist between neighbouring phases along the direction \(\beta\), equivalently like a gauge-field in the \(XY\) model. We next use Eq.~\eqref{eq:action_charged_gamma} with Eq.~\eqref{eq:edge_compressibility_S} to obtain the two contributions to the helicity modulus listed in Eqs.~\eqref{eq:epsilon_theta_alpha} and \eqref{eq:I_theta_alpha} of the main text.

\section{Equivalence between \(\chi_{\Sigma_m}^{\alpha}\) and \(\Upsilon_\alpha^\theta\)}\label{ap:Equivalence}
In this section we show the equivalence between the susceptibility introduced in Eq.~\eqref{eq:susceptibility_theta} and the helicity modulus of \(\theta\). Using Eq.~\eqref{eq:hel_mod_generic} with Eqs.~\eqref{eq:epsilon_theta_alpha} and \eqref{eq:I_theta_alpha}, the explicit expression for the \(\theta\) helicity modulus is 
\begin{equation}\label{eq:hel_mod_theta_full}
    \Upsilon_\alpha^{\theta} = \frac{1}{N}\left[\frac{1}{2V_1}\left\langle\sum_i \cos(\theta_{i+\alpha}-\theta_i)  \right\rangle - \frac{1}{(2V_1)^2}\left\langle\left(  \sum_i \sin(\theta_{i+\alpha} -\theta_i)\right)^2\right\rangle\right].
\end{equation}
Turning to the susceptibility, we start from the correlation function introduced in Eq.~\eqref{eq:susceptibility_theta}
\begin{equation}\label{eq:Correlation_function_sigma_m_2}
    G_{\Sigma_m}^\alpha(r_i - r_j) = \langle \Sigma_\alpha m_i \Sigma_\alpha m_j \rangle -\langle \Sigma_\alpha m_i \rangle \langle \Sigma_\alpha m_j \rangle 
\end{equation}
with 
\begin{equation}
    \Sigma_\alpha m_i = (m_i + m_{i+\alpha})e^{iQ\cdot r_i}
\end{equation}
The two expectation values can be calculating by adding a site-dependent factor to the action in the following manner:
\begin{equation}
    Z = \sum_{\{m_i\}}\int_0^{2\pi} \mathcal{D}\phi e^{-S},
\end{equation}
\begin{equation} \label{eq:action_site_dep_gamma}
        S = i\sum_{i}m_i(\nabla_\tau \phi_i) -J\sum_{i\alpha} \cos(\phi_i - \phi_{i+\alpha}) + V_0 \sum_i m_i^2 + V_1\sum_{i\alpha}(m_i + m_{i+\alpha})^2 
    + i\sum_i (\Sigma_\alpha m_i) \gamma_{i,\alpha}.
\end{equation}
Using the standard expression for calculating expectation values we then have
\begin{equation}\label{eq:Sigma_mi}
    \langle \Sigma_\alpha m_i \rangle = \frac{1}{Z} \sum_{\{m_i\}} \int_0^{2\pi} \mathcal{D}\phi (\Sigma_\alpha m_i) e^{-S}\Big\lvert_{\gamma_{i,\alpha = 0}} = \frac{1}{Z} \sum_{\{m_i\}}\int_0^{2\pi}\mathcal{D}\phi \left(i\frac{\partial}{\partial \gamma_{i,\alpha} }\right)e^{-S}\Big\lvert_{\gamma_{i,\alpha = 0}}
\end{equation}
\begin{equation}\label{eq:Sigma_mimj}
     \langle \Sigma_\alpha m_i \Sigma_\alpha m_j \rangle = \frac{1}{Z} \sum_{\{m_i\}} \int_0^{2\pi} \mathcal{D}\phi (\Sigma_\alpha m_i)(\Sigma_\alpha m_j) e^{-S}\Big\lvert_{\gamma_{i,\alpha = 0}} = \frac{1}{Z} \sum_{\{m_i\}}\int_0^{2\pi}\mathcal{D}\phi \left(i\frac{\partial}{\partial \gamma_{i,\alpha} }\right)\left(i\frac{\partial}{\partial \gamma_{j,\alpha} }\right)e^{-S}\Big\lvert_{\gamma_{i,\alpha = 0}}  
\end{equation}
Comparing Eqs.~\eqref{eq:action_m_with_gamma} and \eqref{eq:action_site_dep_gamma}, the only difference is that \(\gamma_{i,\alpha}\) is site dependent in the latter. This will not change the derivation following \eqref{eq:action_m_with_gamma}, so we can repeat the steps in Appendix \ref{ap:Edge_compressibility} while keeping the site dependence. Thus the partition function takes the form 
\begin{equation}
    Z = \int_0^{2\pi}\mathcal{D}\phi\mathcal{D}\theta e^{-S}
\end{equation}
\begin{equation}
        S = -J\sum_{i\alpha} \cos(\phi_i - \phi_{i+\alpha}) - \frac{1}{2V_1}\sum_{i\alpha}\cos(\theta_i - \theta_{i+\alpha} + \gamma_{i,\alpha}) - \frac{1}{2V_0}\sum_{i}\cos(e^{-iQ\cdot r_i}(\nabla_\tau \phi_i) + \nabla^2 \theta_i).
\end{equation}
Inserting this action into Eqs.~\eqref{eq:Sigma_mi} and \eqref{eq:Sigma_mimj}, we find
\begin{equation}
    \langle\Sigma_\alpha m_i\rangle = i\frac{1}{2V_1}\langle -\sin(\theta_i - \theta_{i+\alpha})\rangle
\end{equation}
\begin{equation}
    \langle \Sigma m_i \Sigma m_j \rangle = \frac{1}{2V_1}\langle\delta_{ij}\cos(\theta_i-\theta_{i+\alpha})\rangle - \frac{1}{(2V_1)^2}\langle\sin(\theta_i-\theta_{i+\alpha})\sin(\theta_j - \theta_{j+\alpha})\rangle
\end{equation}
Finally, inserting this into Eq.~\eqref{eq:Correlation_function_sigma_m} (or Eq.~\eqref{eq:Correlation_function_sigma_m})  and writing out the expression for the susceptibility in Eq.~\eqref{eq:susceptibility_theta} we find
\begin{alignat}{1}
    \chi_{\Sigma_m}^{\alpha} = \lim_{q\to 0} \frac{1}{N^2}\sum_{ij}\Big[ &\frac{1}{2V_1}\langle\delta_{ij}\cos(\theta_i-\theta_{i+\alpha})\rangle - \frac{1}{(2V_1)^2}\langle\sin(\theta_i-\theta_{i+\alpha})\sin(\theta_j - \theta_{j+\alpha})\rangle -\\
    & + \frac{1}{(2V_1)^2}\langle -\sin(\theta_i - \theta_{i+\alpha})\rangle \langle -\sin(\theta_j - \theta_{j+\alpha})\rangle \Big]e^{iq(r_i-r_j)}
\end{alignat}
The final last is current-like, and should be zero. Hence we have neglected it in most of the expressions for the helicity modulus, but it is corresponds to a similar term as the first one in Eq.~\eqref{eq:edge_compressibility_S}. Comparing the two first terms to the expression for the helicity modulus in Eq.~\eqref{eq:hel_mod_theta_full}, we see that the only difference is the order that the long wavelength limit, sum over lattice sites and MC averaging is taken. 

\section{Phase diagram Monte Carlo results}\label{ap:More_MC_results}
In this section we present results for the SC-I transition and I-CSC transition, as representative examples of how we map out the phase diagram in Fig.~\ref{fig:phase_diagram}. Results for the I-CSC transition with \(V_0 = 0.5\) and \(V_1 = 0.8\) are shown in Fig.~\ref{fig:results_PT_I_CSC}. These show a BKT transition in the field \(\phi\) at the critical coupling \(J\simeq 1.03\). Results for the SC-I transition with \(V_0 = 0.5\) and \(J=0.75\) are shown in Fig.~\ref{fig:results_PT_SC_I}. These show a simultaneous \(3D\) \(XY\) transition in \(\phi\) and BKT transition in \(\theta\) at the critical coupling \(V_1 \simeq 0.48\).

\begin{figure}[htb]
    \subfloat[\label{fig:results_PT_I_CSC_a}]{
        \includegraphics[width= 0.3\linewidth]{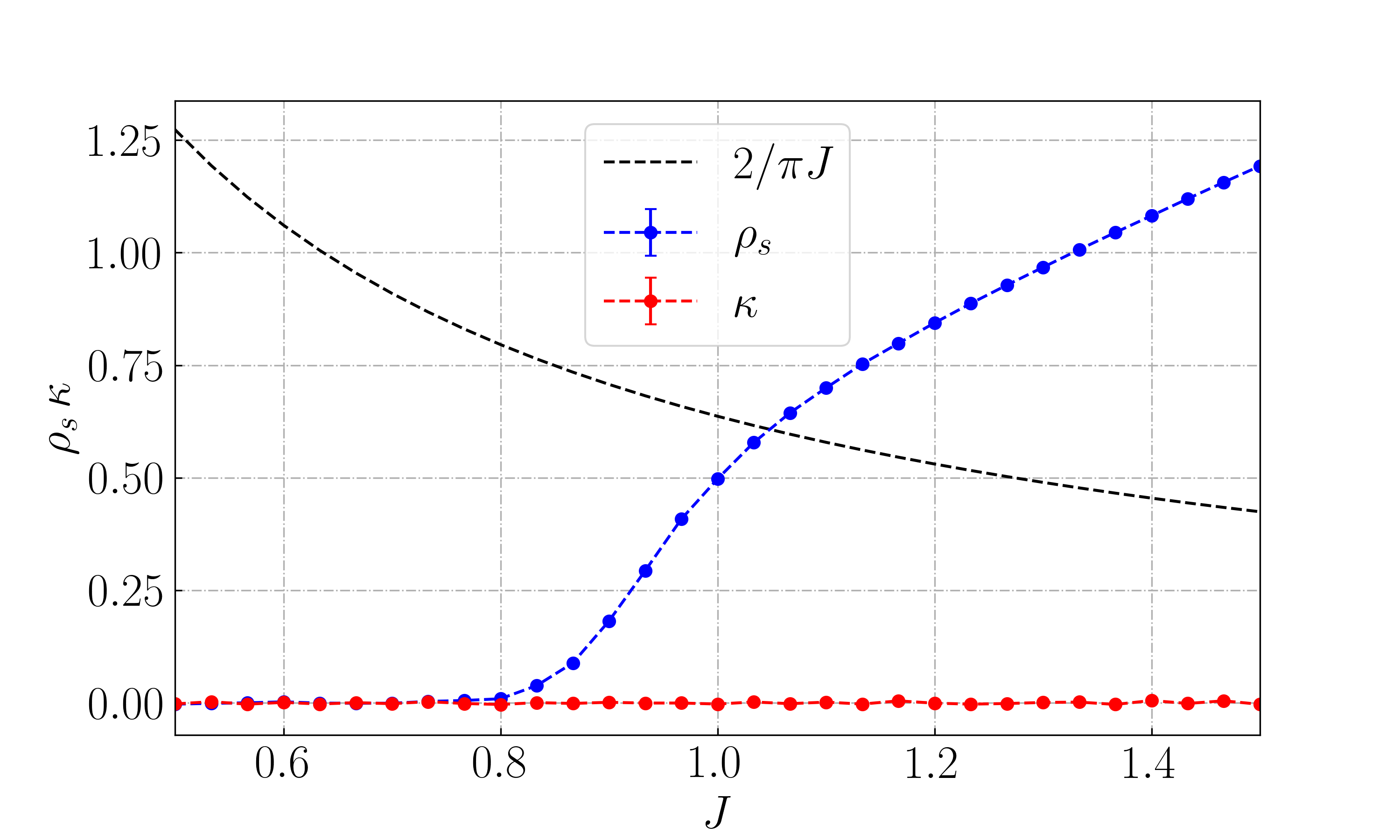}
    } 
    \subfloat[\label{fig:results_PT_I_CSC_b}]{
        \includegraphics[width= 0.3\linewidth]{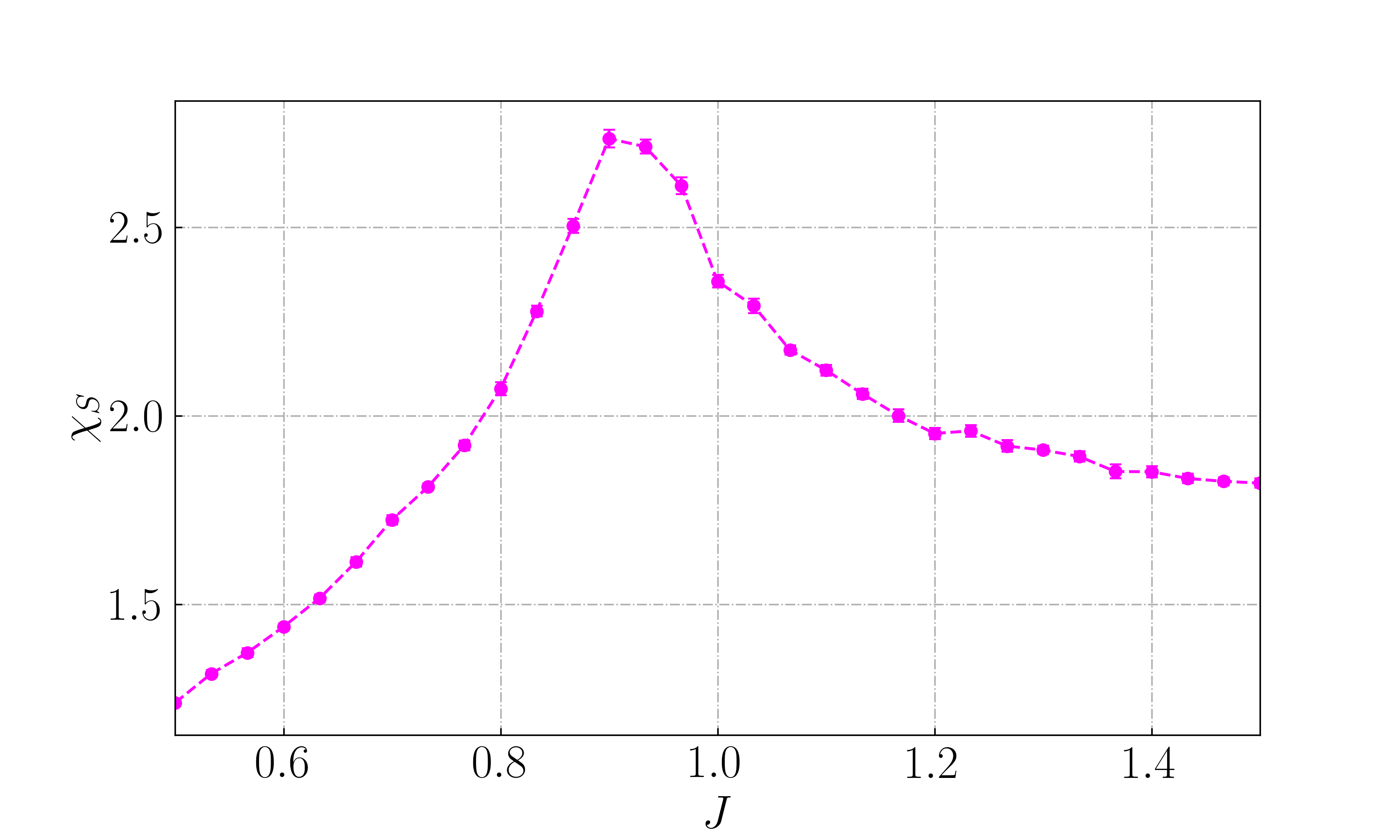}
    } 
    \caption{Results for the I-CSC phase transition, using the model in Eq.~\eqref{eq:model_charge_picture_disc} with \(V_1=0.8\), \(V_0=0.5\) and \(L=L_\tau = 20\). (a) Bulk compressibility and superfluid density given by Eqs.~\eqref{eq:bulk_compressibility}, \eqref{eq:SF_stiffness}. (b) Action susceptibility given by Eq.~\eqref{eq:action_susceptibility}. Shows a BKT transition in \(\phi\) at \(J \simeq 1.03\)}
    \label{fig:results_PT_I_CSC}
\end{figure}

\begin{figure}[htb]
    \subfloat[\label{fig:results_PT_SC_I_a}]{
        \includegraphics[width= 0.3\linewidth]{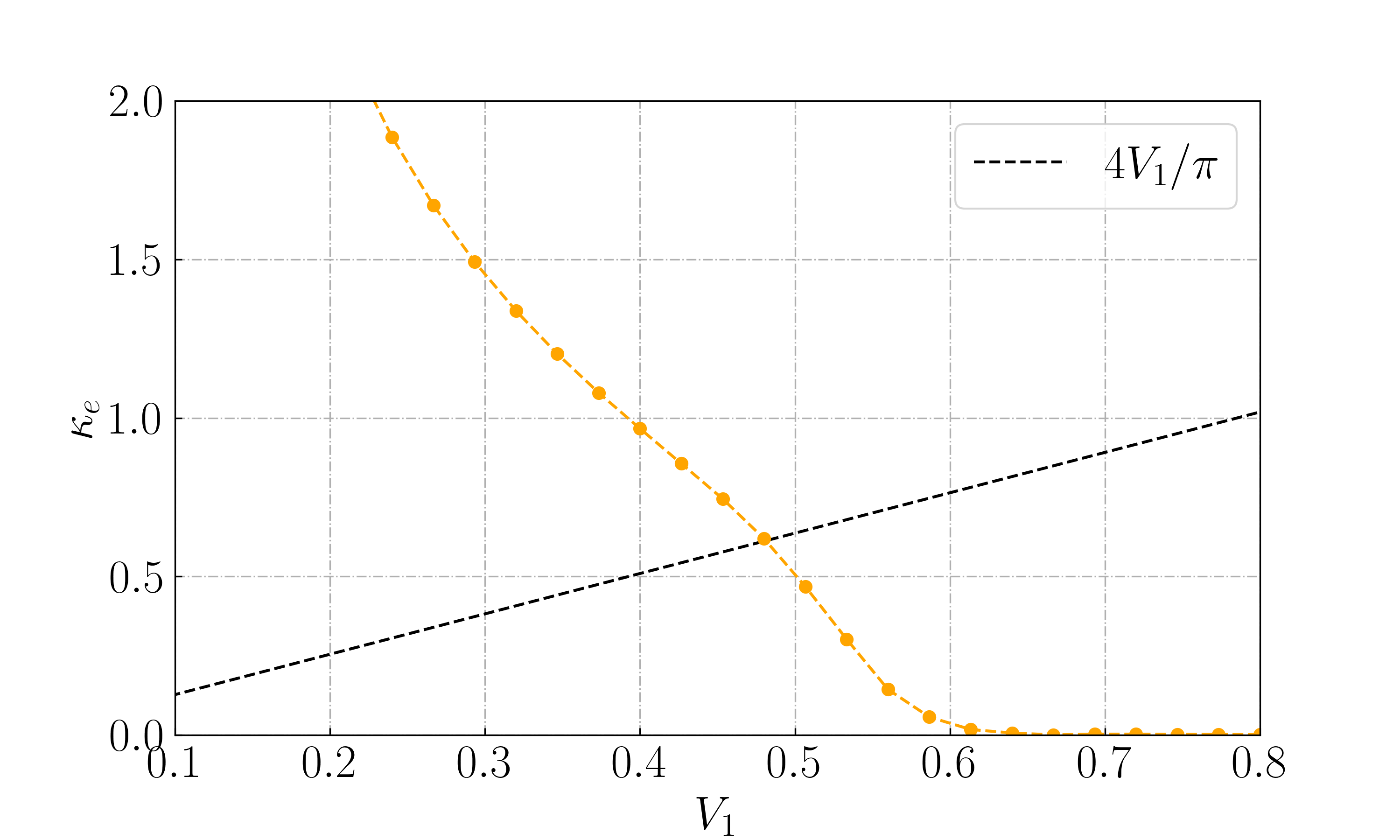}
    } 
    \subfloat[\label{fig:results_PT_SC_I_b}]{
        \includegraphics[width= 0.3\linewidth]{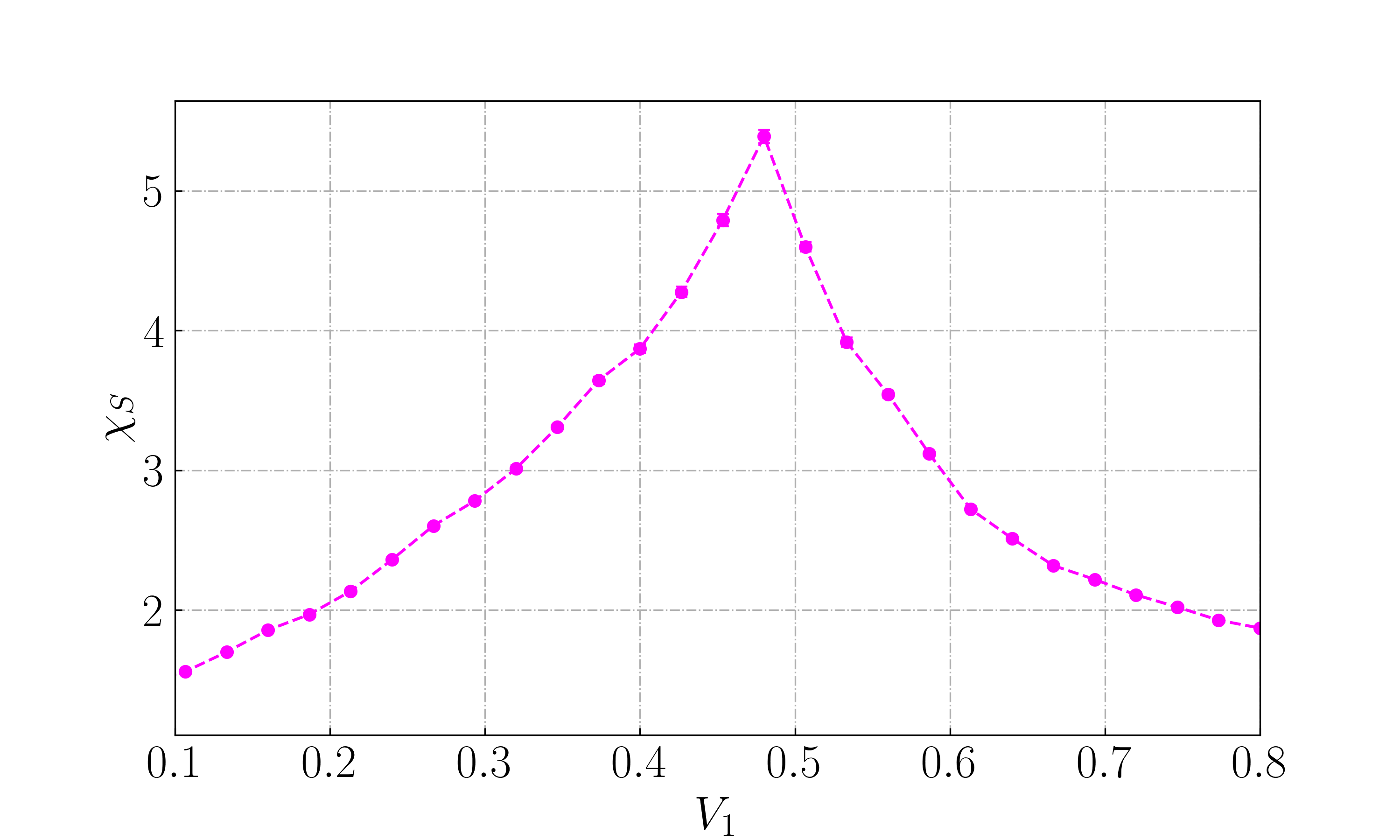}
    } 
    \subfloat[\label{fig:results_PT_SC_I_c}]{
        \includegraphics[width = 0.3\linewidth]{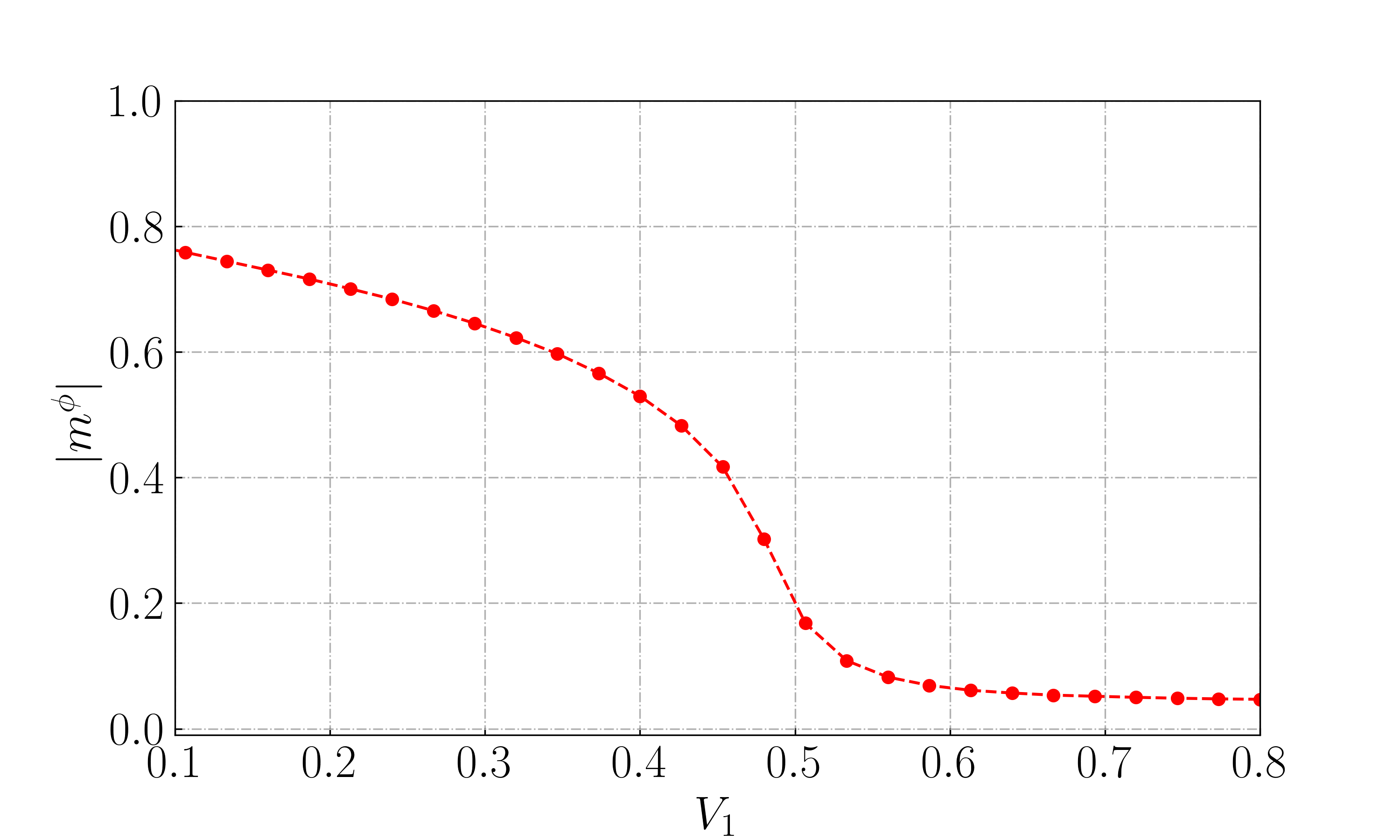}
    }
    \caption{Results for the SC-I phase transition, using the model in Eq.~\eqref{eq:model_charge_picture_disc} with \(J=0.75\), \(V_0=0.5\) and \(L=L_\tau =  20\). (a) Edge compressibility given by Eq.~\eqref{eq:edge_compressibility}. (b) Action susceptibility given by Eq.~\eqref{eq:action_susceptibility}. (c) Order parameter of \(\phi\) given by Eq.~\eqref{eq:OP_phi}. Shows a BKT transition in \(\theta\) and a \(3D\) \(XY\) transition \(\phi\) at \(V_1 \simeq 0.48\).}
    \label{fig:results_PT_SC_I}
\end{figure}

\end{widetext}
\end{document}